\documentclass{aa}

\usepackage[utf8]{inputenc} 
\usepackage[varg]{txfonts}
\usepackage{amssymb}
\usepackage{epsfig}
\usepackage{graphics}
\usepackage{amsmath}
\usepackage{color}
\usepackage{natbib}
\usepackage{hyperref}
\usepackage{gensymb}
\usepackage{bm}
\usepackage{mathtools}
\usepackage{graphicx}
\usepackage{lipsum}
\usepackage{ulem}

\newcommand{\be}{\begin{equation}}
\newcommand{\ee}{\end{equation}}
\def\bc{\begin{center}}
\def\ec{\end{center}}
\def\beq{\begin{eqnarray}}
\def\eeq{\end{eqnarray}}

\newcommand{\msun}{{M}_{\sun}}

\newcommand{\df}[1]{\mathrm{d}{#1}}

\usepackage{physics}

\bibpunct{(}{)}{;}{a}{}{,} 

\makeatletter
\def\fvec#1{\underline{\sbox\tw@{$#1$}\dp\tw@\z@\box\tw@}}
\makeatother

\begin{document}

\setcitestyle{notesep={; }}

\title{Magnetospheric return-current-heated atmospheres of rotation-powered millisecond pulsars}

\titlerunning{Heated atmospheres of rotation-powered millisecond pulsars}

\author{Tuomo~Salmi\inst{1,2}
\and Valery~F.~Suleimanov\inst{3,4,5}
\and Joonas~N\"attil\"a\inst{ 6,7,2}
\and  Juri~Poutanen\inst{1,2,5}}

\institute{Tuorla Observatory, Department of Physics and Astronomy, FI-20014 University of Turku, Finland\\ \email{thjsal@utu.fi, juri.poutanen@utu.fi}
\and Nordita, KTH Royal Institute of Technology and Stockholm University, Roslagstullsbacken 23, SE-10691 Stockholm, Sweden
\and Institut  f\"ur  Astronomie  und  Astrophysik,  Kepler  Center  for  Astro  and  Particle  Physics,  Universit\"at  T\"ubingen,  Sand  1, 72076 T\"ubingen, Germany
\and Astronomy Department, Kazan (Volga region) Federal University, Kremlyovskaya str. 18, 420008 Kazan, Russia
\and Space Research Institute of the Russian Academy of Sciences, Profsoyuznaya str. 84/32, 117997 Moscow, Russia 
\and Physics Department and Columbia Astrophysics Laboratory, Columbia University, 538 West 120th Street, New York, NY 10027, USA
\and Center for Computational Astrophysics, Flatiron Institute, 162 Fifth Avenue, New York, NY 10010, USA
}

\date{Received 26 February 2020 / Accepted 15 June 2020}

\abstract{
We computed accurate atmosphere models of rotation-powered millisecond pulsars in which the polar caps of a neutron star (NS) are externally heated by magnetospheric return currents.
The external ram pressure, energy losses, and stopping depth of the penetrating charged particles were computed self-consistently with the atmosphere model, instead of assuming a simplified deep-heated atmosphere in radiative equilibrium.
We used exact Compton scattering formalism to model the properties of the emergent X-ray radiation.
The deep-heating approximation was found to be valid only if most of the heat originates from ultra-relativistic bombarding particles with Lorentz factors of $\gamma \gtrsim 100$.
In the opposite regime, the atmosphere attains a distinct two-layer structure with an overheated optically thin skin on top of an optically thick cool plasma.
The overheated skin strongly modifies the emergent radiation: it produces a Compton-upscattered high-energy tail in the spectrum and alters the radiation beaming pattern from limb darkening to limb brightening for emitted hard X-rays.
This kind of drastic change in the emission properties can have a significant impact on the inferred NS pulse profile parameters as performed, for example, by Neutron star Interior Composition ExploreR.  
Finally, the connection between the energy distribution of the return current particles and the atmosphere emission properties offers a new tool to probe the exact physics of pulsar magnetospheres.
}
 
\keywords{radiative transfer -- methods: numerical -- pulsars: general -- stars: atmosphere -- stars: neutron -- X-rays: stars}

\maketitle
 
\section{Introduction}
\label{sec:intro}

Rotation-powered millisecond pulsars (RMPs), also known as radio millisecond pulsars, are rapidly rotating neutron stars (NSs), which are believed to have been spun-up due to the accretion from a companion star \citep{RS82,ACR82}. 
After transiting from the accretion-powered to the rotation-powered phase, the NS polar caps are, as opposed to accretion, heated by the bombardment of electrons and positrons from a magnetospheric return current \citep[see e.g.][]{ZShak69,AW73,RS75,ZTZ95}. 
The rotation of the resulting hotspots gives rise to X-ray pulsations. 

Pulse profiles of millisecond pulsars carry information about the NS mass and radius \citep{PG03,PB06,MLC07,ML15}. 
These pulses can be modelled using a general relativistic ray-tracing framework \citep[e.g.][]{PB06,MLC07,NP18,SNP18} to constrain the unknown equation of state (EOS) of cold dense matter \citep{lattimer12,OF16,WAC16,WYP19}.
In accreting millisecond pulsars, the largest uncertainty in modelling the pulse profiles comes from the poorly known emission pattern of the hotspots \citep{PG03} as well as from the variability of the pulse profiles \citep[e.g.][]{IP09}. 
In this respect, the RMPs are potentially better targets, as their pulse profiles are stable on the time scales of years and therefore can be accurately measured, for instance, with the Neutron star Interior Composition ExploreR (NICER) mission \citep{nicer2016}.
Their emission pattern was also expected to be well-described by a hydrogen atmosphere model \citep{bogdanov2013,bogdanov2016}, which has also been used recently to model the NICER data \citep{BLM_nicer19,MLD_nicer19,RWB_nicer19}.
However, the exact spectral energy distribution and the angular emission pattern of radiation emitted by the hotspots affect the pulse shape, and they are therefore important to be accurately modelled \citep{SNP18}. 
They depend on the details of the return-current particle distribution and the resulting atmosphere structure, which are currently unknown \citep{BPO19}.

Atmospheres of the NSs have been previously modelled in many different studies.
Using a plane-parallel atmosphere model for RMPs in radiative equilibrium, the angular and energy distribution of the radiation has been described, for example, by \citet{bogdanov2013} \citep[see also e.g.][]{ZPS96,HRN06,HH09,mcphac2012}.
A similar model, but with an exact formulation for Compton scattering, is studied in \citet{SSP19}.
However, the assumption of the radiative equilibrium is only valid if the in-falling pair plasma heating the polar caps deposits its energy at very deep layers.
It is therefore necessary to study the additional physical processes for atmospheres heated, in their different layers, by a beam of bombarding particles.
Atmospheres heated from the top have previously been modelled by \citet{GCZT19} for strongly magnetised NSs (in the grey approximation) and for accreting NSs by \citet{DDS2001} and \citet{SPW18}. 
In the case of RMPs, the effects of stopping the magnetospheric particles have been studied by \citet{BPO19}, although using a simplified atmosphere model.
In this work, we aim to study the return-current-heating effects using the full atmosphere model including the exact treatment of Compton scattering, which is expected to be important for the energy balance of the heated outer layers of the atmosphere and for the observed spectrum, in particular, its angular distribution.

The remainder of this paper is structured as follows.
In Sect.~\ref{sec:model}, we explain the theoretical framework, including how we modelled the return-current-heated NS atmosphere, emphasising the differences to the existing models. 
In Sect.~\ref{sec:results}, we calculate NS atmospheres for a set of model parameters, especially for different energies or energy distributions of the bombarding particles, and compare the results with each other and to the non-heated model.
We discuss the reasons for the differences and implications for the NICER parameter constraints in Sect.~\ref{sec:discussion}.
We conclude in Sect.~\ref{sec:conclusions}.

\section{The model}\label{sec:model}
\subsection{Main equations}\label{sec:main_eqs}

We consider steady-state plane-parallel atmospheres with additional heating in the surface layers caused by the magnetospheric return current.
According to \citet{arons81} and \citet{HM02}, we could expect that the particles penetrating NS atmosphere would primarily be positrons.
However, we still consider the loss rates only for electrons, because they lose their energy in the same way to positrons and have a simpler form of energy-loss equations \citep{Berger1984,BPO19}.
We also assume the NS atmosphere to consist purely of hydrogen, which is justified by a short gravitational stratification timescale \citep{AI1980,ZP2002},
leaving only the lightest elements to the layers which determine the properties of the escaping radiation.

The code that we use in atmosphere modelling is a modified version from \citet{NSK15}, which is based on \citet{SPW12} and \citet{K70}.
Our model has a set of equations very similar to those presented in \citet{SPW18}.
However, instead of considering the energy loss of accreted protons, we consider the case for relativistic return current electrons. 
The heating rate of the NS atmosphere is expressed in terms of relative luminosity $l = F/F_{\mathrm{Edd}} = T_{\mathrm{eff}}^{4}/T_{\mathrm{Edd}}^{4}$, where
\be\label{eq:fedd_def}
F_{\mathrm{Edd}} = \frac{gc}{\kappa_{\rm e}} = \sigma_{\mathrm{SB}} T_{\mathrm{Edd}}^4 
\ee
is the Eddington flux and $T_{\mathrm{Edd}}$ is the corresponding Eddington temperature, which is the maximum effective temperature on the NS surface for which local flux does not exceed the Eddington one for the given NS surface gravitational acceleration $g$ (a free parameter of the model). Here $\sigma_{\mathrm{SB}}$ is the Stefan-Boltzmann constant, and the Eddington flux is evaluated using Thomson scattering opacity $\kappa_{\rm e} =\sigma_{\mathrm{T}}/m_{\mathrm{p}} \approx 0.4\,\mbox{cm}^2\ \mbox{g}^{-1}$, 
where $\sigma_{\mathrm{T}}$ is the Thomson cross-section and $m_{\mathrm{p}}$ is the proton mass.

The emergent bolometric flux is obtained as a sum of two components,
\be\label{eq:emerg_bol_flux}
F = \sigma_{\mathrm{SB}}(T_{\mathrm{eff,i}}^{4}+T_{\mathrm{eff,h}}^{4}) = \sigma_{\mathrm{SB}}T_{\mathrm{Edd}}^{4}(l_{\mathrm{i}}+l_{\mathrm{h}})  = F_{\mathrm{i},0}+F_{\mathrm{h},0},
\ee
where the lower indices $\rm i$ and $\rm h$ refer to the intrinsic NS flux and that added by  return-current-heating, respectively.
In all of our models, we put $l_{\mathrm{i}} = l_{\mathrm{h}}/100$.
The exact value is not expected to be important as long as $l_{\mathrm{h}}$ is much greater than $l_{\mathrm{i}}$.
The number flux (per unit area) of bombarding electrons is related to the heating rate as 
\be\label{eq:m_acc}
\dot{N}_{\rm e} = \frac{F_{\mathrm{h},0}}{m_{\mathrm{e}} c^2 
(\langle\gamma_{\mathrm{0}}\rangle-1)}\ \mbox{cm}^{-2}\mbox{s}^{-1},
\ee
where $\langle\gamma_{\mathrm{0}}\rangle$ is the average Lorentz factor of the in-falling particles
and $m_{\mathrm{e}}$ is the electron (positron) mass.

The computation follows otherwise \citet{SPW12}, but the absorption opacities and the electron number densities are obtained directly from the Kramers' opacity law and the ideal gas law, since we assume a fully ionised hydrogen atmosphere with free-free opacity as the only source of absorption opacity.
The contribution from return-current-heating is added in the hydrostatic equilibrium equation which reads as 
\be\label{eq:hdbal}
\frac{1}{\rho}\frac{\df P}{\df r} = -g-g_{\mathrm{ram}}+g_{\mathrm{rad}},
\ee
where $P$ is the gas pressure, $g_{\mathrm{rad}}$ is the radiative acceleration (defined in \citealt{SPW12}),  and $g_{\mathrm{ram}}$ is the additional ram pressure acceleration \citep{SPW18}. 
For monoenergetic in-falling particles the ram acceleration is  
\be\label{eq:gram}
g_{\mathrm{ram}} = - \dot{N}_{\rm e} m_{\mathrm{e}} \dv{}{m} (\gamma v) =-\dot{N}_{\rm e} m_{\mathrm{e}}c \frac{\gamma}{p}\frac{\df \gamma}{\df m} , 
\ee
where $\df m = -\rho \df r$ is the column density, $\rho$ is the plasma density of the atmosphere, $v= \beta c$ is the velocity of the particles, $\gamma$ is the corresponding Lorentz factor, and $p=\sqrt{\gamma^2-1}$ is the dimensionless momentum at the given position $m$. 

The energy balance equation is written as 
\be\label{eq:ene_balance}
2\pi\int_{0}^{\infty}\df x\int_{-1}^{+1} [\sigma(x,\mu)+k(x)][I(x,\mu)-S(x,\mu)]\df \mu = -Q^{+},
\ee
where $k(x)$ is the ``true'' absorption opacity, and $\sigma(x,\mu)$ is the electron scattering opacity. 
The general form of the local energy dissipation rate $Q^{+}$ is \citep{SPW18}:
\be\label{eq:qbig}
Q^{+} = \dot{N}_{\rm e} m_{\mathrm{e}}  q^{+} = - \dot{N}_{\rm e} m_{\mathrm{e}} c^2 \dv{}{m} (\gamma - 1) =  - \dot{N}_{\rm e} m_{\mathrm{e}} c^2  \dv{\gamma}{m},
\ee
where $q^{+}$ is the specific particle deceleration. 

The dependence of particle deceleration on depth $\df \gamma / \df m$ fully determines the atmosphere model.
To solve it, we can compute the energy loss rate along the radial direction $r$ (which we assume to be the same as the displacement direction because the particles penetrate very closely to the normal direction) as
\be
\label{eq:dedt_displ}
\dv{E}{t}  = - \beta c \dv{E}{r} = - \beta c \dv{(m_{\mathrm{e}}c^{2}(\gamma -1))}{r} 
= \rho m_{\mathrm{e}}c^{3}\ \beta \dv{\gamma}{m} .
\ee
Another rather general expression for the energy losses can be written as 
\be \label{eq:en_loss_gen} 
\dv{E}{t}  = - n_{\mathrm{e}} \sigma_{\mathrm{T}} m_{\mathrm{e}}c^{3} 
\beta \Phi(\gamma),
\ee
where $\Phi$ is an effective cross-section in units of the Thomson cross-section, and $n_{\mathrm{e}}$ is the number density of the electrons in the atmosphere. 
These two equations then transform to the following differential equation for the Lorentz factor 
\be \label{eq:lorentz_loss} 
\dv{\gamma}{\tau}  = -  \Phi(\gamma) ,
\ee
where $\df \tau=\kappa_{\mathrm{e}} \df m$ is the Thomson optical depth. 
This differential equation can be solved numerically to obtain $\gamma(m)$ with a given $T(m)$, $\rho(m)$, and $n_{\mathrm{e}}(m)$.  
The solution is then used to determine the local energy dissipation rate and additional acceleration due to ram pressure.
We note that the solution $\gamma(m)$ is largely independent of the atmospheric structure, except for a residual dependence of the cross-section $ \Phi(\gamma)$ on density  \citep{Berger1984}. 
The dependence of the cross-section on the particle Lorentz factor is considered in the next section.

\subsection{Stopping of fast electrons}\label{sec:stop_elec}

In a highly or moderately relativistic case, the energy loss of the electrons takes place via the bremsstrahlung radiation (dominating above $500$ MeV or $\gamma \approx 1000$), due to the Coulomb collisions with particles in the plasma, and due to excitation of collective modes of plasma (Langmuir waves). 
However, it could be that electrons with energies much larger than the thermal energy of the atmosphere electrons cannot excite Langmuir waves effectively.
The total energy loss cross-section $\Phi (\gamma)$ (in Eq. \eqref{eq:lorentz_loss}) is obtained by summing the cross-sections for all the different mechanisms.

An accurate approximation at all energies for the effective cross-section of electron-proton bremsstrahlung derived from the analytic solutions of \citet{Heitler1954} using Born approximation in the non-relativistic and extreme relativistic limits is given by \citet{H04}: 
\begin{eqnarray}\label{ep_cross_brems}
\Phi_{\mathrm{rad}}^{\mathrm{e^{-}p}} (\gamma) &=& \frac{3\alpha}{\pi} \frac{ \gamma^{3}}{p^2+\gamma^{2}}
\left\{\frac{\ln(\gamma + p)}{\beta} - 
\frac{1}{3} \right.
\nonumber \\
&+&\left. \frac{p^2}{\gamma^{6}} \left[ \frac{2}{9} \gamma^{2} 
- \frac{19}{675} \gamma p^2 -0.06\frac{p^4}{\gamma} \right] \right\},
\end{eqnarray}
where $\alpha$ is the fine structure constant. 
For the electron-electron bremsstrahlung cross-section, we use different analytic approximations at different electron energies, that are also presented in \citet{H04}.
For example, in case of ultra-relativistic energies ($\gamma > 1000$) we have 
\be \label{eq:ee_cross_brems_ultra}
\Phi_{\mathrm{rad}}^{\mathrm{e^{-}e^{-}}}(\gamma)  \approx \alpha \frac{3}{2\pi} (\gamma-1) \left[ \ln (2\gamma) - \frac{1}{3}\right].
\ee
 
The energy loss of the electron due to the combined effect of Coulomb collisions with electrons (assuming that the electron velocity $v$ is much larger than the thermal velocity $v_{\mathrm{th}}$ of the plasma electrons at that depth) and Langmuir waves, is given by \citet[][see also \citealt{Gould1972}]{SB2008}: 
\begin{eqnarray}\label{eq:dedt_coloumvb_elec}
\Phi_{\mathrm{C+L}}  (\gamma) & = & \frac{3} {4\beta^2}
\left[ \ln \left( \frac{(\gamma -1) \beta^{2}} {2\epsilon_{\mathrm{p}}^2 } \right)+ 1 \right. \nonumber \\
&+&\left. \frac{1}{8}\left(\frac{\gamma-1}{\gamma} \right)^{2}
-\left(\frac{2\gamma-1}{\gamma^{2}} \right)\ln2  \right],
\end{eqnarray}
where $\epsilon_{\mathrm{p}}=\hbar \omega_{\mathrm{p}}/m_{\mathrm{e}} c^2$,  $\omega_{\mathrm{p}}=\sqrt{4\pi n_{\mathrm{e}} e^{2}/m_{\mathrm{e}}}$ is the plasma frequency.
The effect of electron-proton Coulomb collisions can be neglected because the energy loss is inversely proportional to the mass of the target. 

\subsection{Return current with energy distribution}
\label{sec:eqs_for_multiene}

In case the in-falling electrons have some energy distribution, the total energy deposition rate is obtained by integrating the product of local dissipation rate and the distribution function of the particles over the initial Lorentz factor $\gamma_{0}$ as 
\be\label{eq:intg_fdist_mom}
Q^{+}(m) =  \dot{N}_{\rm e} m_{\rm e} \int_{\gamma_{\mathrm{min}}}^{\gamma_{\mathrm{max}}} f(\gamma_{0}) q^{+}(\gamma_{0},m) \df \gamma_{0},
\ee
where we set the maximum to $10^{5}$ MeV ($\gamma_{\mathrm{max}} \approx 2\times 10^{5}$), and the minimum $\gamma_{\mathrm{min}}$  is our free parameter.
We use a power-law distribution of the return-current particle number over the Lorentz factor as suggested by observations of gamma-ray pulsars and simulations of pulsar magnetospheres \citep{HM2001,CPS2016,BKT2018}:  
\be\label{eq:beskin_formula}
f(\gamma_{0}) = N \gamma_{0}^{-\delta},
\ee
where $N = 1/\int_{\gamma_{\mathrm{min}}}^{\gamma_{\mathrm{max}}}\gamma_{0}^{-\delta} \df \gamma_0$ is the normalisation constant, and the exponent $\delta$ is a free parameter in our model. 
The average Lorentz factor for the incoming return-current beam is given as (if $\delta \neq 2$)
\be \label{eq:gamma_avg1}
\langle \gamma_0 \rangle =  \int_{\gamma_{\mathrm{min}}}^{\gamma_{\mathrm{max}}} N \gamma_0^{1-\delta} \mathrm{d}\gamma_0 = \frac{N(\gamma_{\mathrm{max}}^{2-\delta}-\gamma_{\mathrm{min}}^{2-\delta})}{2-\delta}. 
\ee
If $\delta = 2$, we have
\be \label{eq:gamma_avg2}
\langle \gamma_0 \rangle =  N \ln(\frac{\gamma_{\mathrm{max}}}{\gamma_{\mathrm{min}}}).
\ee

For the energy deposition rate, we get now
\be\label{eq:big_Qa_final}
Q^{+}(m) = 
-\dot{N}_{\mathrm{e}} \ m_{\mathrm{e}}c^2  \int_{\gamma_{\mathrm{min}}}^{\gamma_{\mathrm{max}}} f(\gamma_0)
\frac{\df \gamma}{\df m} \df \gamma_{0}, 
\ee
where $\df\gamma/\df m$ is a function of $\gamma_{0}$ and $m$.
Integration over the column density should give us the total energy deposition rate per unit area: 
\be \label{eq:number_flux}
\int_0^\infty Q^{+}(m) \df m = F_{\rm h,0} = 
\dot{N}_{\rm e} (\langle \gamma_0 \rangle -1)  m_{\rm e} c^2,
\ee 
which determines the number flux of electrons $\dot{N}_{\mathrm{e}}$ appearing in Eq.\,\eqref{eq:big_Qa_final}. 

The expression for the additional acceleration due to ram pressure created by the penetrating particles (used in the hydrostatic balance equation) becomes \citep[see again][]{SPW18}
\begin{align}
\label{eq:g_acc_final}
g_{\mathrm{ram}} = 
-\dot{N}_{\mathrm{e}} \ m_{\mathrm{e}}c 
\int_{\gamma_{\mathrm{min}}}^{\gamma_{\mathrm{max}}} f(\gamma_0) 
\frac{\df (\gamma\beta) }{\df m}   
\df \gamma_{0} .
\end{align}

\subsection{Computation of the atmosphere structure}\label{sec:comp_of_atmos}

We computed the model atmospheres using similar setup in energy and optical depth grids as in \citet{SPW12}.
The formal solution for radiation transfer equation was found using the short-characteristic method \citep{OK87} first in three angles, but finally with eleven angles in the end of the temperature iterations.
The parabolic approximation of the solution was replaced with the linear approximation to avoid negative intensities.
The full solution was found using the accelerated $\Lambda$-iteration method described in \citet{SPW12}.

\begin{figure}
\centering
\resizebox{\hsize}{!}{\includegraphics{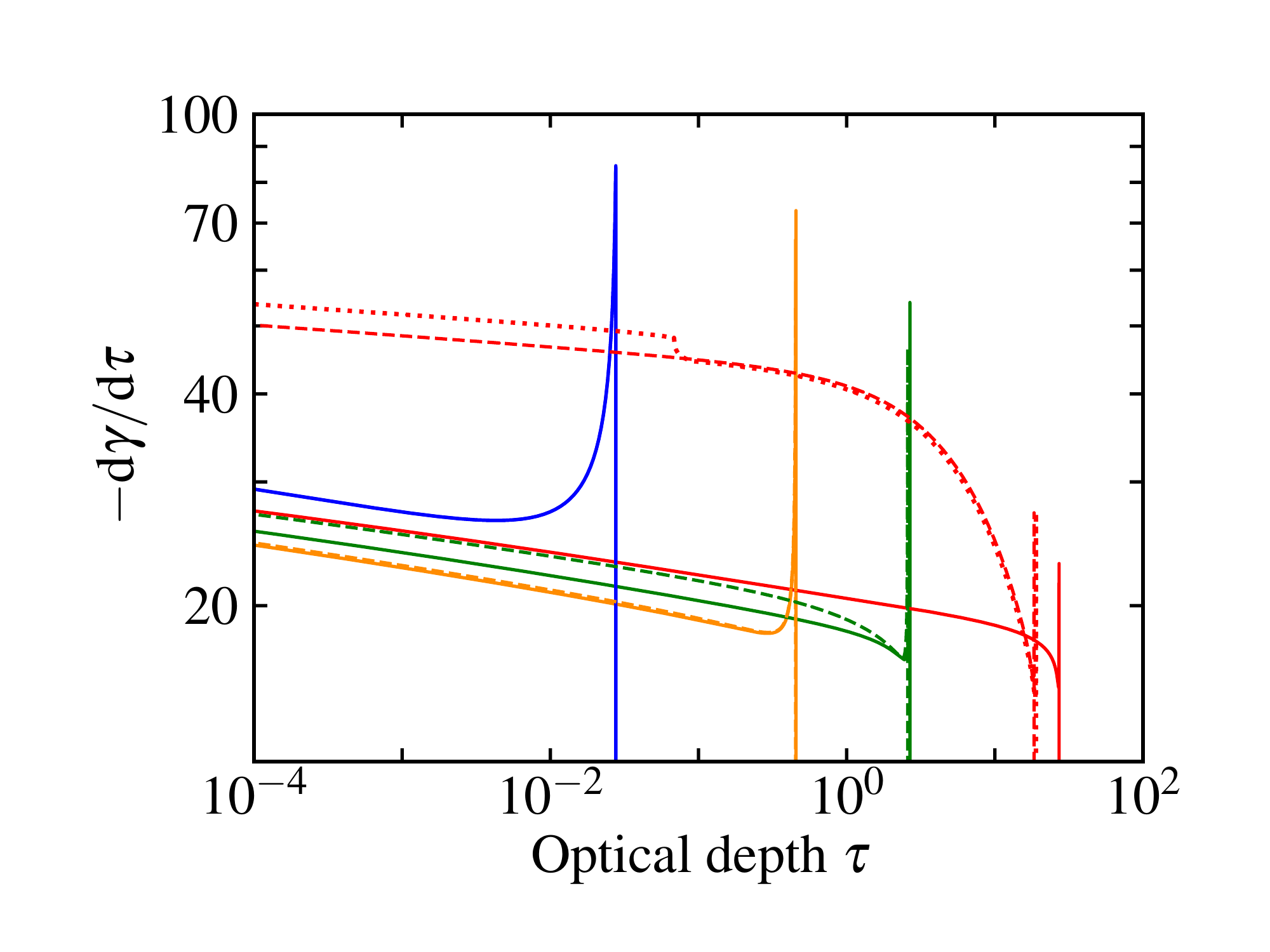}} 
\caption{Electron deceleration as a function of Thomson optical depth for different values of incoming beam Lorentz factor $\gamma_{0}$ for grey atmosphere. 
Blue, orange, green, and red curves correspond to energy losses due to Coulomb collisions and wave excitation for $\gamma_{0} = 2, 10, 50$, and $500$, respectively. 
The dashed lines show the same for energy losses where also bremsstrahlung is taken into account.
The dotted line shows the result for $\gamma_{0} = 500$ when we use a return-current-heated atmosphere instead of grey atmosphere (for model parameters given in Sect. \ref{sec:comp_eloss}).
Inclusion of bremsstrahlung losses is found to be important for particles with $\gamma_0 \gtrsim 100$.  
}
\label{fig:comp_col_brems}
\end{figure}

Compared to \citet{SPW12}, the temperature corrections were modified so that the relative flux error (used in Avrett-Krook flux correction) was 
\be\label{eq:rel_flux_error}
\epsilon_{\mathrm{F}} (m) = 1-\frac{F_{\mathrm{i},0}+ F_{\mathrm{h},0}-\int_{0}^{m}Q^{+}(m')\df m'}{ \int_{0}^{\infty}F_{\mathrm{x}}(m)\df x },  
\ee
where $x$ is the photon energy.
In addition, the energy balance error, used in the temperature corrections for upper atmospheric layers, became 
\begin{eqnarray} \label{eq:ene_bal_error}
\epsilon_{\lambda} (m)\!\!\! & =&\!\!\! \frac{w}{2}\int_{0}^{\infty}\!\!\!\df x \int_{-1}^{+1}\!\!\![\sigma(x,\mu)+k(x)][I(x,\mu)-S(x,\mu)]\df \mu  \nonumber \\ &+& wQ^{+}(m), 
\end{eqnarray}
where $w$ is a weight used to adjust the correction at the beginning of the iterations.

The energy dissipation rate $Q^{+}(m)$ was calculated for each iteration using Eq.\,\eqref{eq:big_Qa_final} for a distribution of particles or Eq.\,\eqref{eq:qbig} for mono-energetic particles.
When solving  Eq.\,\eqref{eq:lorentz_loss} to find $\gamma (m)$  we assumed that the bombarding electrons have lost all their energy at the depth where their velocity drops below $\gamma = 1.1$ (as mentioned above, the equations given in Sect.\,\ref{sec:stop_elec} are only valid for fast electrons).  

We performed our model computations by starting from a model atmosphere with equal intrinsic and return-current-heated temperatures, $T_{\mathrm{eff,i}} = T_{\mathrm{eff,h}}$.
During the first tens of iterations, we applied only the flux correction in order to have the innermost part of the atmosphere converged.
After that, we steadily increased $T_{\mathrm{eff,h}}$ up to its final value, and started linearly increasing the weight, $w$, on the energy balance error  
(and decreasing the maximum allowed temperature correction from flux error)
when making temperature corrections. 
Typically, during the last hundreds of iterations, only the outer layers of the NS were heated until reaching the energy balance.
 
\section{Results}\label{sec:results}

\subsection{Comparison of energy loss mechanisms}\label{sec:comp_eloss}

Let us first discuss the details of the particle stopping and energy loss mechanisms.
We began by computing the deceleration of the return current particles as a function of the Thomson optical depth using a simplified atmosphere model (grey atmosphere) to confirm that the results are similar to those of \citet{BPO19}.
We also studied, when the inclusion of the bremsstrahlung energy losses (neglected in all the previous works) becomes important. 
The comparison of electron deceleration is shown in Fig.  \ref{fig:comp_col_brems}, in case of a grey atmosphere, with $\log g =  14.0$, $T_{\mathrm{eff}}= 4.64$ MK (0.4 keV), and assuming mono-energetic return current particles with $\gamma_{0} = 2$, $10$, $50$, or $500$.

In addition, for $\gamma_{0} = 500$ we present the result using a return-current-heated atmosphere from Sect. \ref{sec:res_multilgg} (with $\delta=3$, $\gamma_{\mathrm{min}}=10$, $T_{\mathrm{eff,h}} = 5$~MK, $T_{\mathrm{eff,i}} = 1.6$~MK, and $\log g = 14.3$). 
This demonstrates that $\df \gamma / \df \tau$ depends only slightly on the atmosphere model.
It only depends on $n_{\mathrm{e}}$ through the plasma frequency in Eq. \eqref{eq:dedt_coloumvb_elec}, and therefore the heated atmosphere with much more rarefied upper layers (structure discussed more later in Sects. \ref{sec:results} and \ref{sec:discussion}) causes slightly faster deceleration.
Our calculations also indicate that the bremsstrahlung losses are important for stopping electrons that have initial Lorentz factors higher than $\gamma_{0} \approx 100$.

\begin{figure}
\centering
\resizebox{\hsize}{!}{\includegraphics{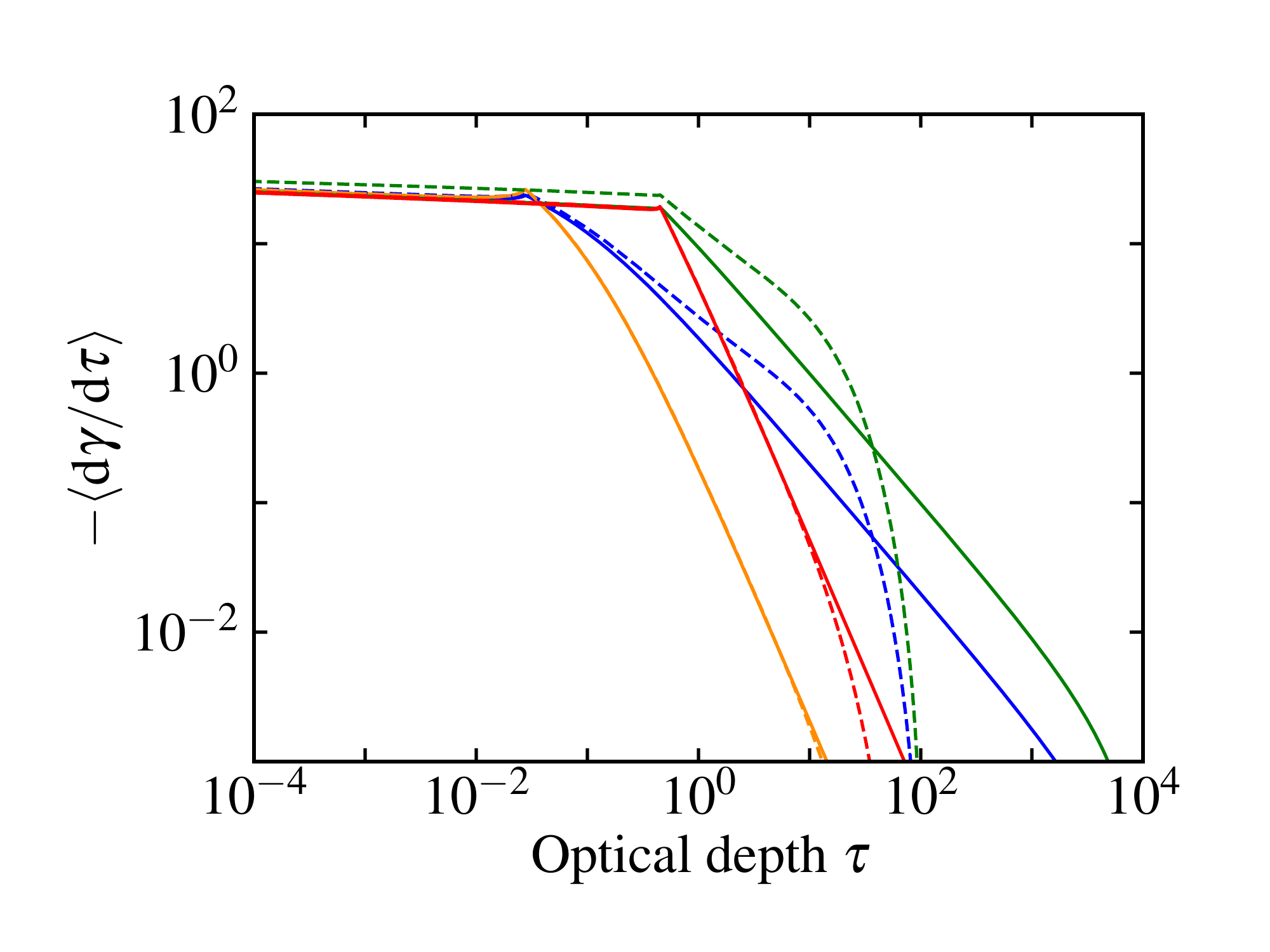}}
\caption{
Comparison of average electron deceleration (see the definition in Sect. \ref{sec:comp_eloss}) for multi-energetic particles using grey atmosphere with different power-law indices $\delta$ and minimum cutoff energies $\gamma_{\mathrm{min}}$ (cf. Fig.~5 in \citealt{BPO19}).
The green curve is for $\delta = 2$ and $\gamma_{\mathrm{min}} = 10$, the blue curve for $\delta = 2$ and $\gamma_{\mathrm{min}} = 2$, the red curve for $\delta = 3$ and $\gamma_{\mathrm{min}} = 10$, and the orange curve for $\delta = 3$ and $\gamma_{\mathrm{min}} = 2$; in all cases the upper limit for the energy distribution is $\gamma_{\mathrm{max}} = 2 \times 10^{5}$.
The dashed curves present the results where the bremsstrahlung losses are taken into account. 
}
\label{fig:eloss_multiene}
\end{figure}

Similarly to \citet{BPO19}, we also detect a sharp peak at the effective stopping depth of the particle.
This is caused by the rapid increase of the energy deposition rate at relatively low particle velocities.
The height of the peak is determined by our choice of the velocity where the particles are considered to have lost all of their energy (see Sect. \ref{sec:comp_of_atmos}).
The equations presented in Sect.\,\ref{sec:stop_elec} are not valid for velocities comparable to thermal velocities of the plasma electrons because the cross-sections diverge at $\beta=0$.
In any case, the peak is not important to our calculation because its contribution to the energy loss is minor; the peak is also not resolved when interpolating the calculated energy losses to the more coarse optical depth grid, where the radiative transfer and temperature corrections were calculated. 
Similar selection of the grid parameters was also used in \citet{SPW12}. 
This coarse grid is used in all the subsequent figures after Fig.\,\ref{fig:eloss_multiene}.

\begin{figure*}
\centering
\includegraphics[width=8cm]{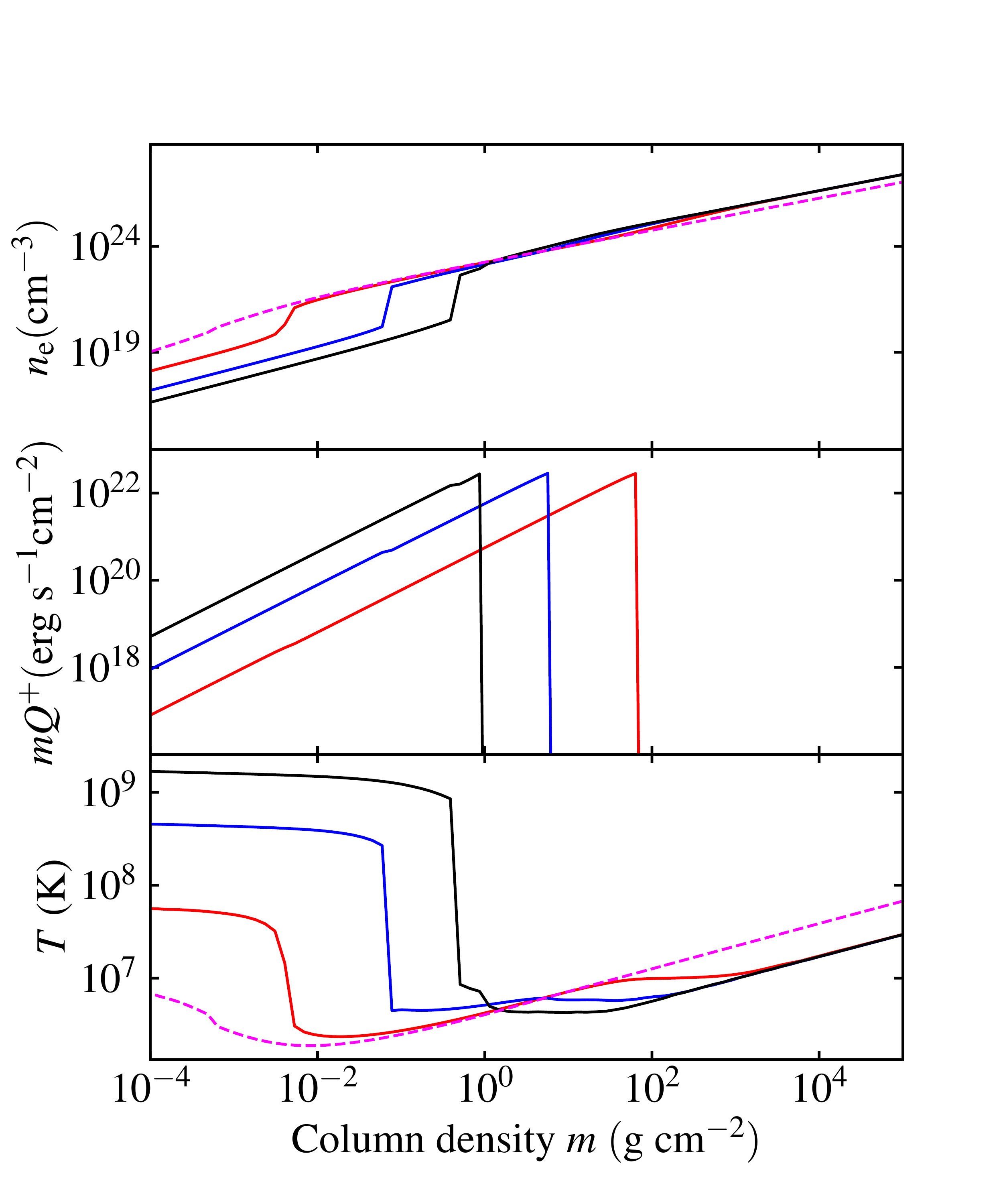} 
\includegraphics[width=8cm]{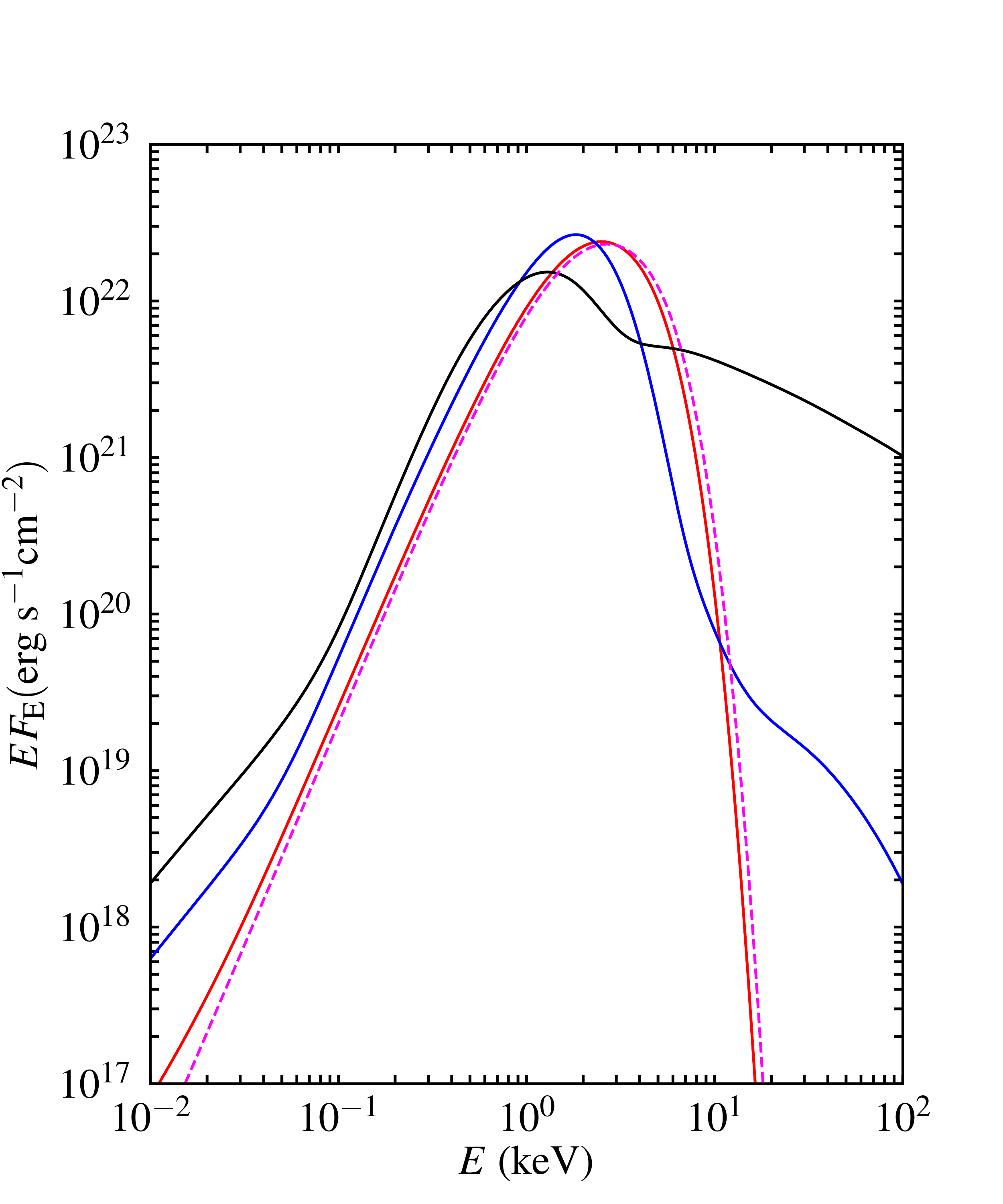} 
\caption{Atmosphere structure (\textit{left}) and the spectrum of the escaping radiation (\textit{right}) for mono-energetic return current. 
\textit{Left panels}: dependence of the electron number density $n_{\mathrm{e}}$, energy loss of the return current $mQ^+$, and the temperature $T$ on the column density. 
The red curves correspond to a model with Lorentz factor $\gamma_{0} = 500$, the blue curves are for $\gamma_{0} = 50$, and the black curves are for $\gamma_{0} = 10$. 
The parameters are $T_{\mathrm{eff,h}} = 5$~MK, $T_{\mathrm{eff,i}} = 1.6$~MK, and $\log g = 14.3$.
The corresponding structure and the spectrum for model without external heating ($T_{\mathrm{eff,i}} = 5$~MK, $T_{\mathrm{eff,h}} = 0$) are shown with the magenta dashed curve.
} 
\label{fig:struc_gall}
\end{figure*}

The average electron deceleration rates for multi-energetic particles (see Sect. \ref{sec:eqs_for_multiene} for more details) and for grey atmosphere are shown in Fig.~\ref{fig:eloss_multiene}, where $\langle \df \gamma / \df \tau \rangle = \int_{\gamma_{\mathrm{min}}}^{\gamma_{\mathrm{max}}}f(\gamma_{0})(\df \gamma / \df \tau) \df \gamma_{0}$. 
In this case, the electrons lose their energy and stop at significantly lower depths if the bremsstrahlung energy loss is taken into account.
We also note that we have a significantly higher deceleration rate (and thus energy loss) in the deeper layers than in \citet[][where the energy deposition rates are cut off around $\tau \approx 10$]{BPO19}. 
This could be an outcome of a different choice for $\gamma_{\mathrm{max}}$ when integrating over the energy distribution. 
In the electron deceleration shown in Fig.~\ref{fig:eloss_multiene}, we detect a cutoff at high depths only if the bremsstrahlung losses are taken into account.
We also note that we made the stopping of the particle less abrupt by using an exponential function to decrease the energy loss rate when the velocity becomes smaller.
This decreases the numerical noise in the results when integrating over different particle Lorentz factors, $\gamma_{0}$ (due to the sharp peaks seen in Fig. \ref{fig:comp_col_brems}) and simultaneously avoids the need to have an extremely fine grid resolution for $\gamma_{0}$.

\subsection{Atmospheres heated by mono-energetic particles}
\label{sec:res_monoene}

Let us now discuss the emergent radiation from atmospheres heated with mono-energetic particles.
In this and in the following sections, we calculate atmosphere models with $T_{\mathrm{eff,h}} = 5.0$~MK, $T_{\mathrm{eff,i}} = 1.6$~MK, and $\log g = 14.3$, unless otherwise stated.
These values are feasible for millisecond pulsars, which are our primary targets.
The effective temperature was chosen roughly correspond to the hottest observed millisecond pulsars \citep{GKR_nicer19} in order to emphasise the effects of external heating (see also Sect. \ref{sec:res_multiT} for different temperatures).
For simplicity, the bremsstrahlung energy losses were not taken into account in the case of mono-energetic particles, because the effect is very small for the chosen particle energies. 
The results of taking the effect into account are only shown in Sect. \ref{sec:res_multiene}, where the particle distributions extend to higher energies.

We began by comparing the spectra and temperature structures from atmosphere models where return-current-heating dominates to that which includes only the intrinsic heat from NS (with $T_{\mathrm{eff,i}} = 5$~MK). 
We also compared the results for mono-energetic particles of different initial energies. 
The results are shown in Fig.~\ref{fig:struc_gall}.
The calculated atmospheres attain a two-layer structure:
the beam of bombarding particles flows first through a hot and rarefied over-heated outer layer and then stops in a cooler and denser inner layer. The over-heated layer has such high temperature because the energy, which is released in this rarefied plasma, cannot be emitted by thermal radiation, and only Compton down-scattering can cool the plasma. 
The last process is effective at high temperatures only. 

For high energy of the in-falling particles, the energy deposition happens in very deep layers and the resulting atmosphere structure resembles closely the model with deep-heated atmosphere in a radiative equilibrium.
The difference in temperature at high depths is closely related to the fact that the intrinsic NS flux in the two models is very different (the effective temperature is three times smaller in heated atmosphere models) and $T\propto T_{\rm eff,i} \tau^{1/4}$. 
Close to the surface layers, we see a temperature inversion where the temperature jumps to $T\approx (6-7)\times10^7$~K (see red curves in left panels of Fig.~\ref{fig:struc_gall}). 
The discontinuity in the electron number density is formed at the same depth.

The penetrating particles that have the lowest energies, deposit most of their energy in the upper layers, therefore the models with $\gamma_0=10$ produce a very hot skin, where $T\approx10^9$~K, and the depth where the temperature inversion occurs increases (see black curves in Fig.~\ref{fig:struc_gall}) compared to the case of high-energy incoming particles. 
This model has largest deviation from a non-heated atmosphere.

\begin{figure*}
\centering
\includegraphics[width=8cm]{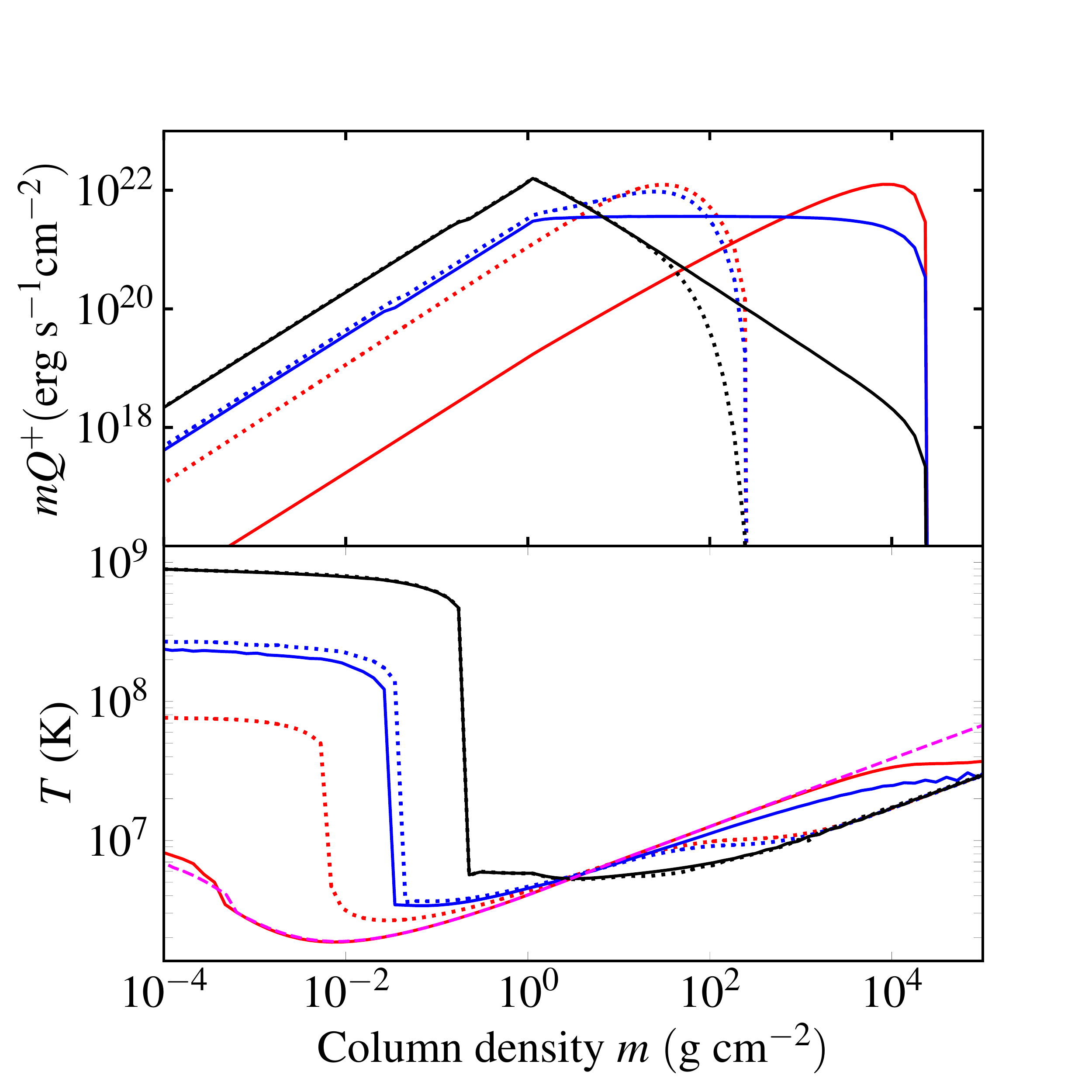} 
\includegraphics[width=8cm]{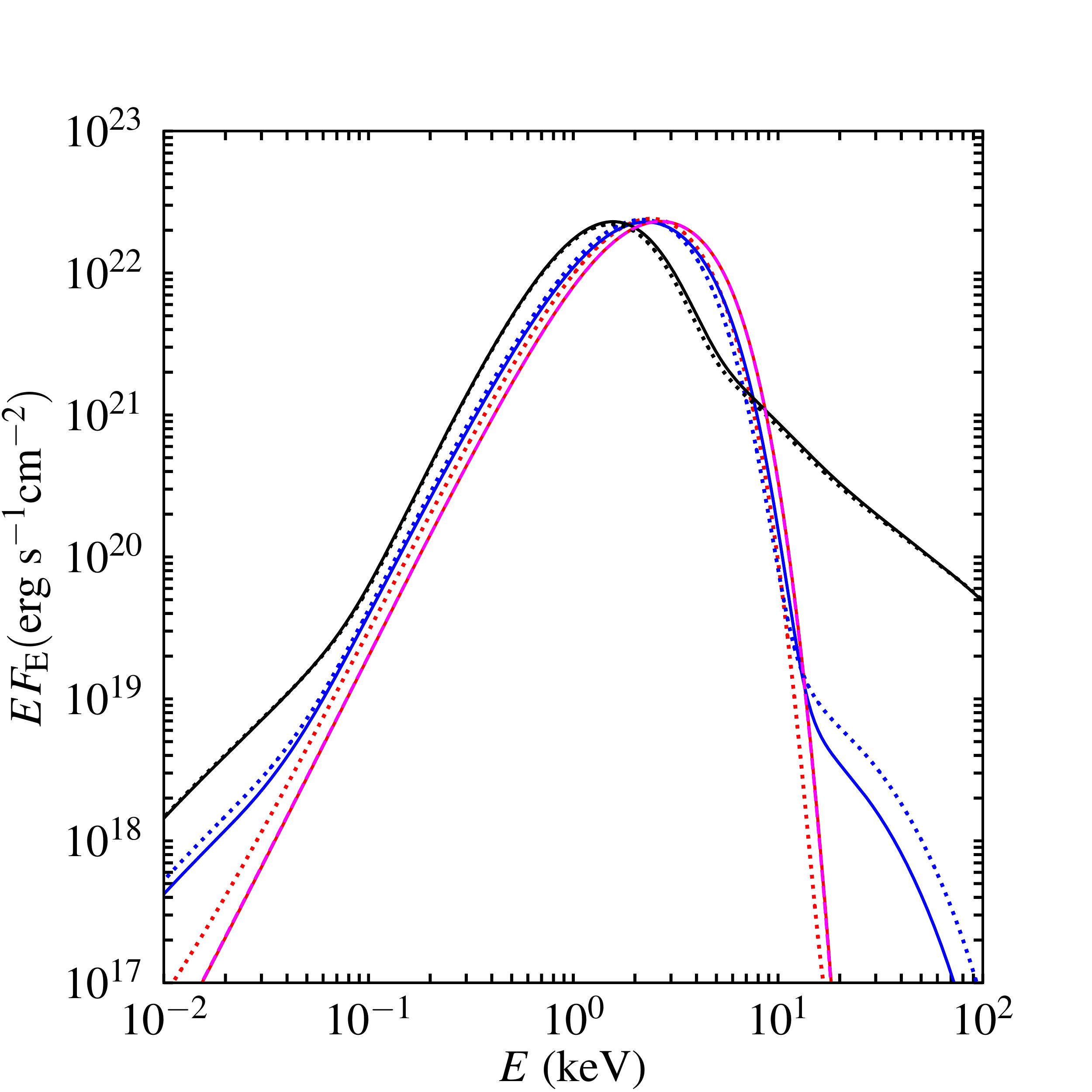} 
\caption{Atmosphere structure (\textit{left}) and the spectrum of escaping radiation (\textit{right}) for a power-law distribution of bombarding particles. 
\textit{Left panels}: energy loss of the return current $mQ^+$ and the temperature $T$ dependence on the column density. 
The red curves correspond to $\delta = 1$, 
the blue curves to $\delta = 2$, and the black curves to $\delta = 3$. 
Other parameters are given in Table \ref{table:params}.
The dotted curves show the corresponding results when also the bremsstrahlung energy losses are taken into account. 
The dashed magenta curves correspond to the temperature structure and escaping spectrum of the atmosphere without external heating (almost completely overlapping the red solid curve). 
}
\label{fig:struc_multiene}
\end{figure*}

\begin{figure*}
\centering
\includegraphics[width=8cm]{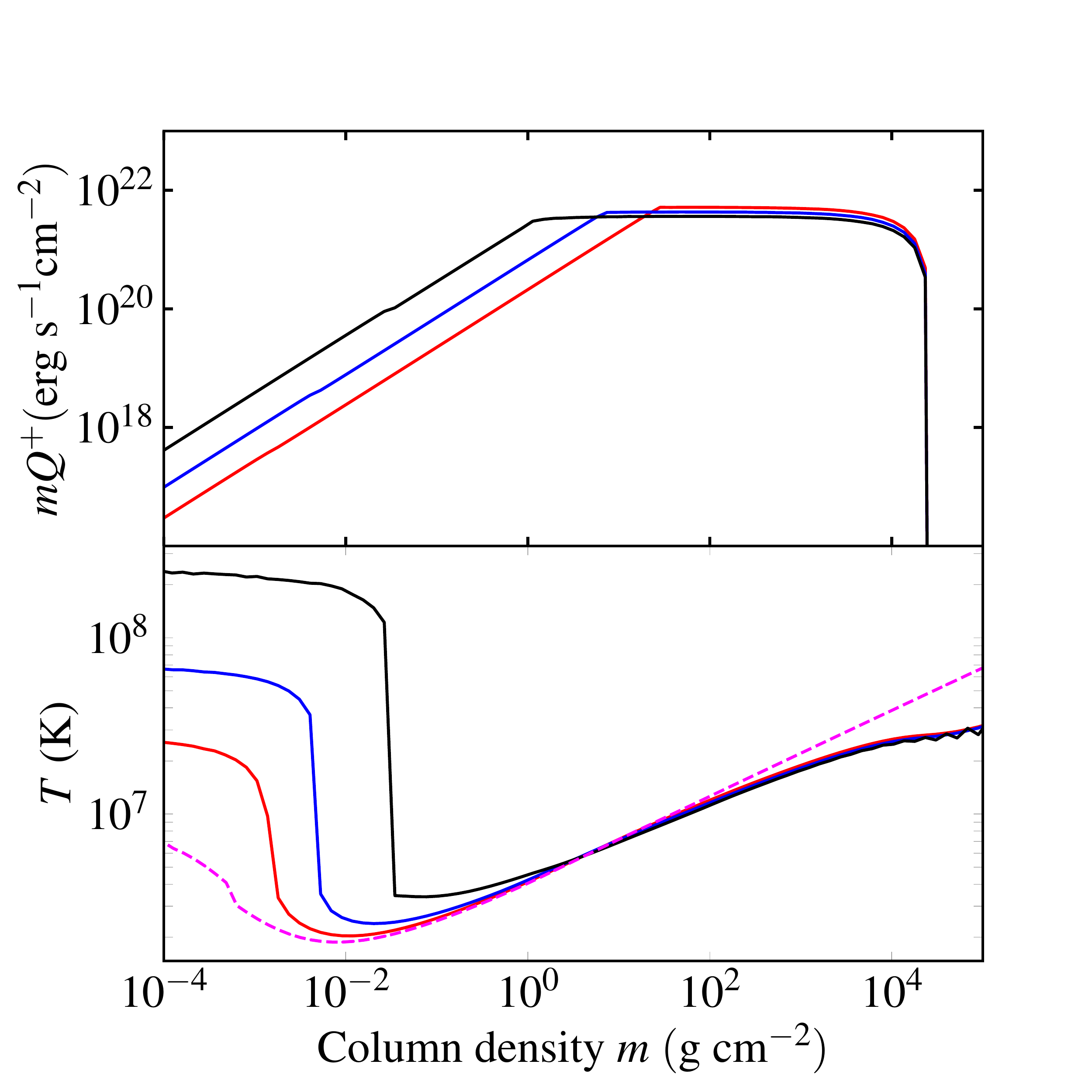} 
\includegraphics[width=8cm]{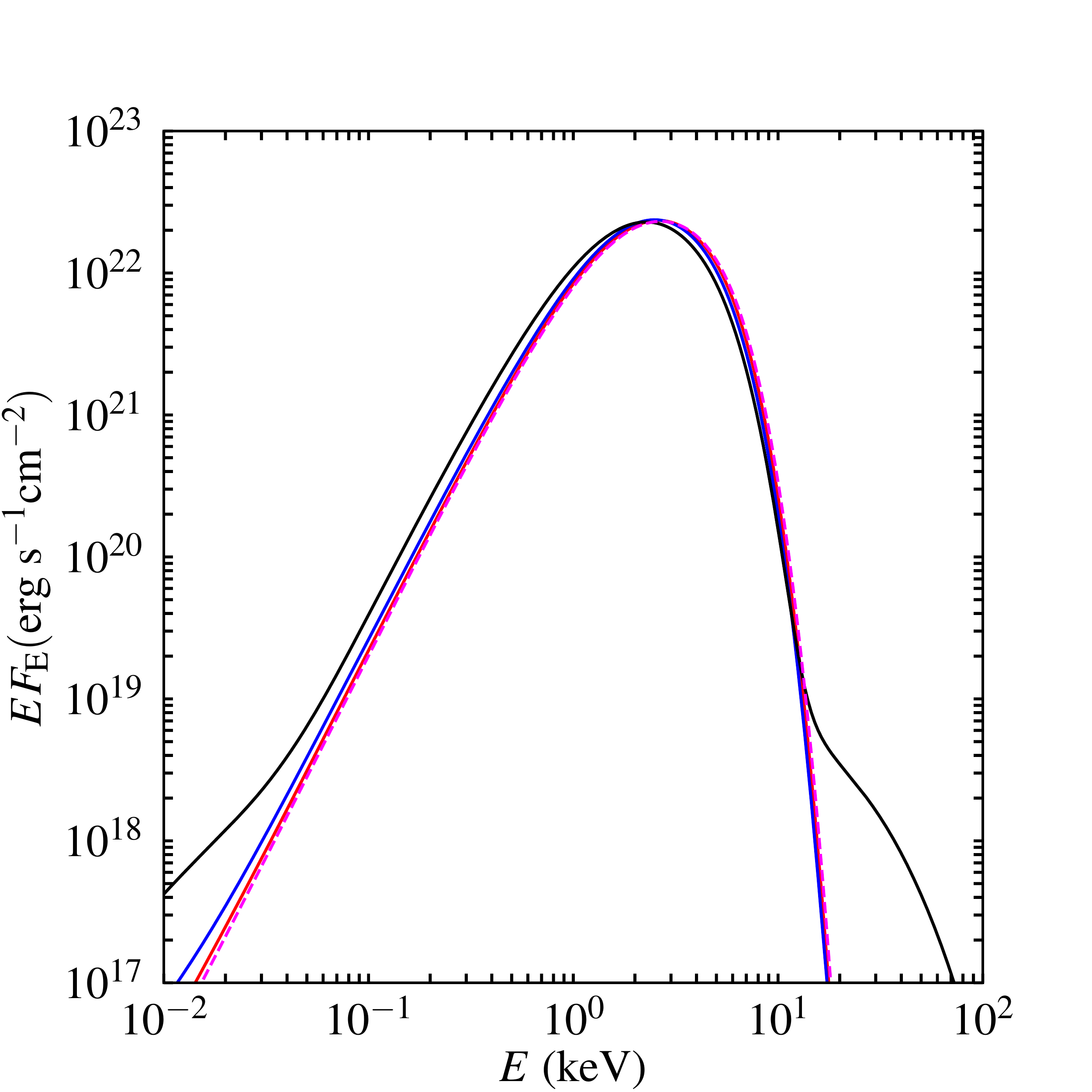} 
\caption{
Atmosphere structure (\textit{left}) and the spectrum of escaping radiation (\textit{right}) for a power-law distribution of bombarding particles with $\delta=2$ for different $\gamma_{\mathrm{min}} = 10$ (black curves), $\gamma_{\mathrm{min}} = 50$ (blue), and $\gamma_{\mathrm{min}} = 200$ (red). 
Other NS atmosphere parameters are given in Table \ref{table:params}. 
The temperature structure and the spectrum without heating are shown with magenta dashed curves. 
}
\label{fig:struc_multigmin}
\end{figure*}

The spectra of the escaping radiation, also presented in Fig. \ref{fig:struc_gall}, show that the models with higher contribution of low-energy particles have their spectral peak at lower energies (especially with $\gamma_{0} = 10$). 
They also have both a high-energy tail and increased emission at low photon energies, compared to the almost black-body like spectra in case of high-energy penetrating particles. 
The spectrum with $\gamma_{0} = 500$ is very similar to the deep-heating model, whereas the spectrum with $\gamma_{0} = 10$ deviates highly from them.
The model with $\gamma_{0} = 50$ shows similar discrepancies when comparing to the deep-heating model, but with a slightly smaller magnitude.

\begin{table}
  \caption{Parameters of the fiducial atmosphere model.} 
\label{table:params}
\centering
  \begin{tabular}[c]{ l  c } 
    \hline\hline
     Parameter & Value 
     \\ 
     \hline 
      Heated effective temperature $T_{\mathrm{eff,h}}$ & 5 MK  \\ 
Ratio of heated to intrinsic luminosity  $l_{\mathrm{h}}/l_{\mathrm{i}}$ & 100  \\ 
      Surface gravity $\log g$ & 14.3 \\ 
      Minimum Lorentz factor $\gamma_{\mathrm{min}}$ & 10  \\ 
      Maximum Lorentz factor $\gamma_{\mathrm{max}}$ & $2\times10^{5}$ \\ 
      Slope of the distribution $\delta$ & 2  \\ 
    \hline
  \end{tabular}
      \tablefoot{The quantities $l_{\mathrm{h}}/l_{\mathrm{i}}$ and $\gamma_{\mathrm{max}}$ are fixed in all of the models. Other parameters are varied in simulations presented in Figs. \ref{fig:struc_multiene}, \ref{fig:struc_multigmin}, \ref{fig:struc_multilgg}, and \ref{fig:struc_multiT}.} 
  \end{table}

\subsection{Atmospheres heated by distribution of particles}
\label{sec:res_multiene}

\begin{figure*}
\centering
\includegraphics[width=8cm]{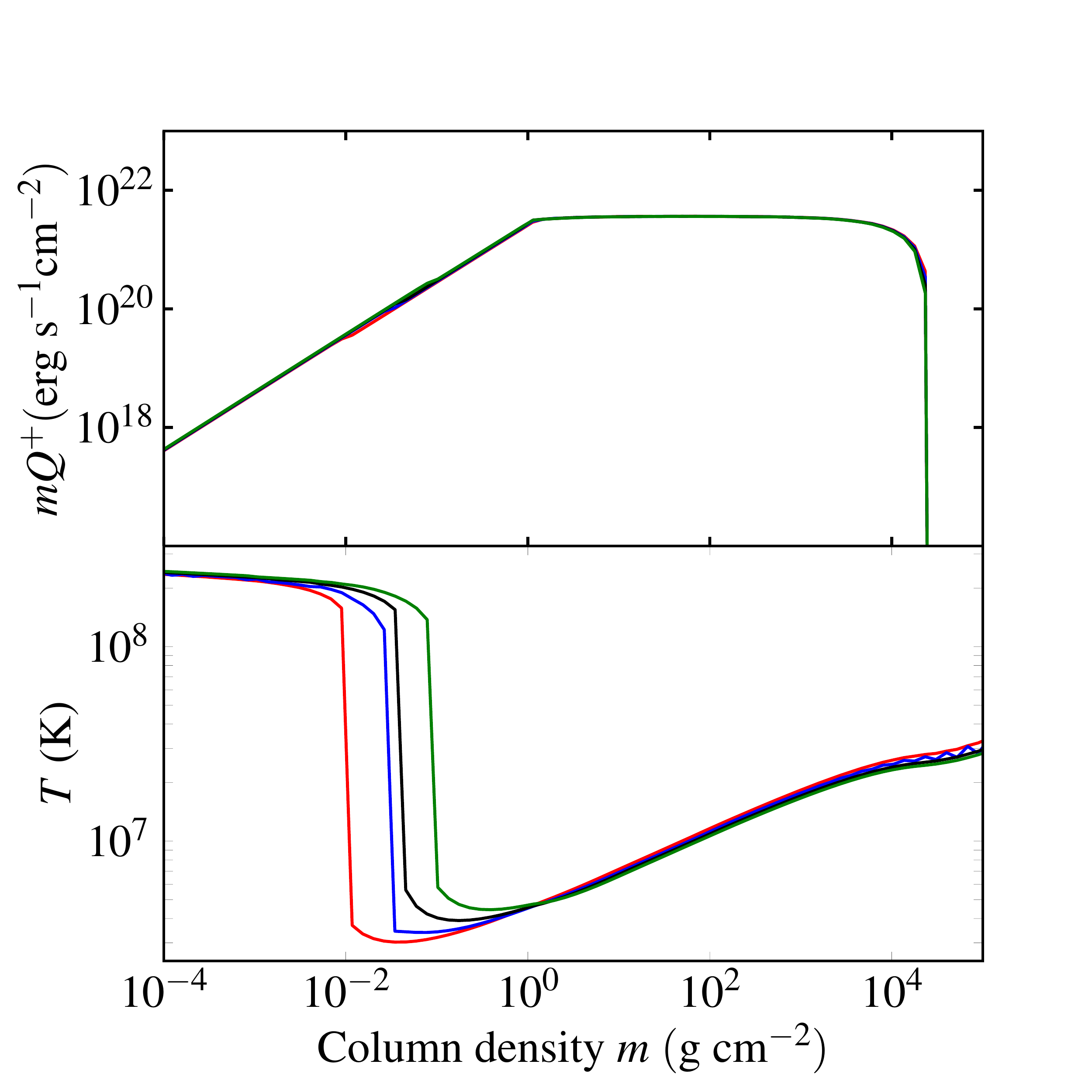} 
\includegraphics[width=8cm]{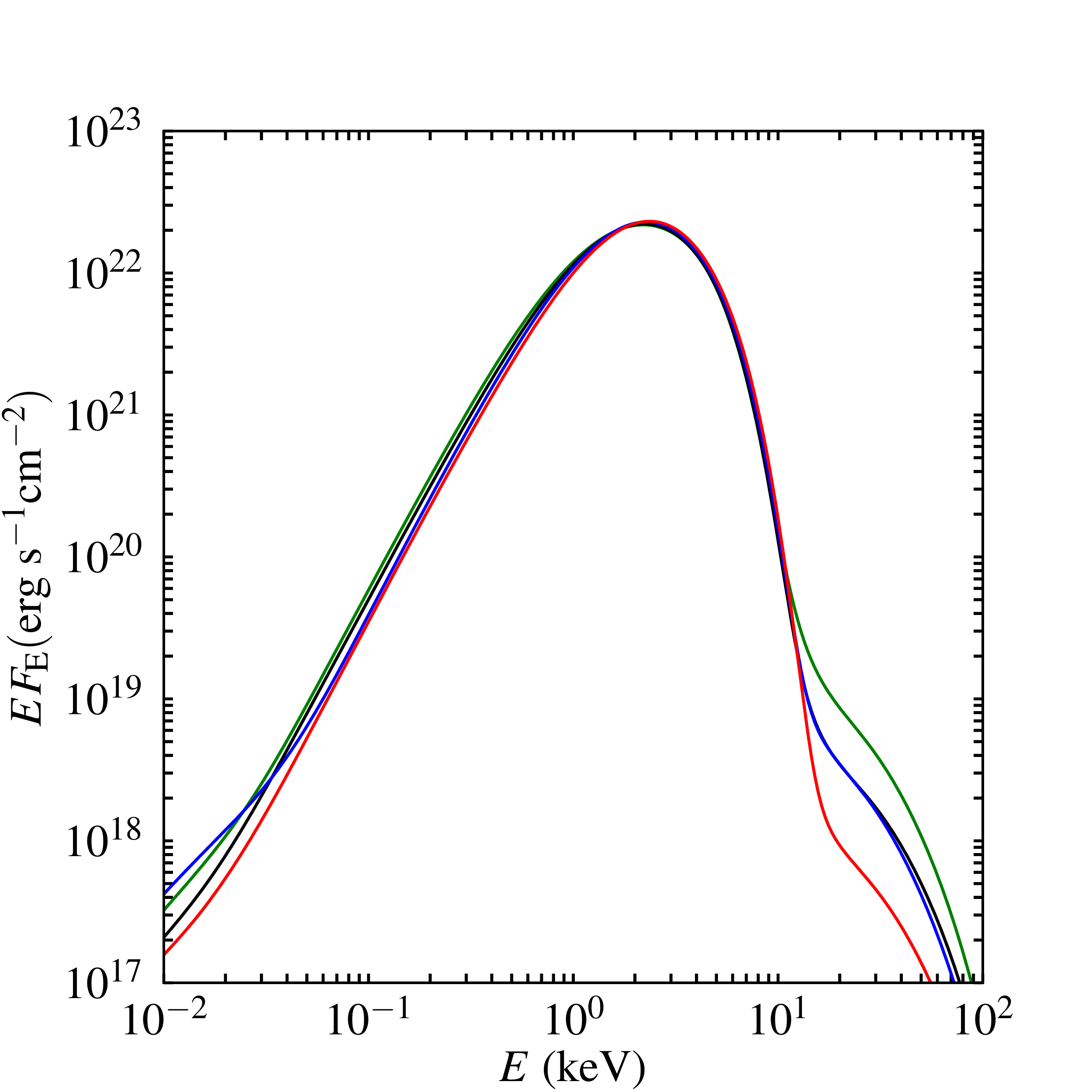} 
\caption{Atmosphere structure (\textit{left}) and the spectrum of escaping radiation (\textit{right}) for a power-law distribution of bombarding particles with $\delta = 2$ and $\gamma_{\mathrm{min}} = 10$ for different NS surface gravities: $\log g = 13.7$ (green curves), 14.0 (black), 14.3 (blue), and 14.6 (red).
Other NS atmosphere parameters are given in Table \ref{table:params}. 
}
\label{fig:struc_multilgg}
\end{figure*}

\begin{figure*}
\centering
\includegraphics[width=8cm]{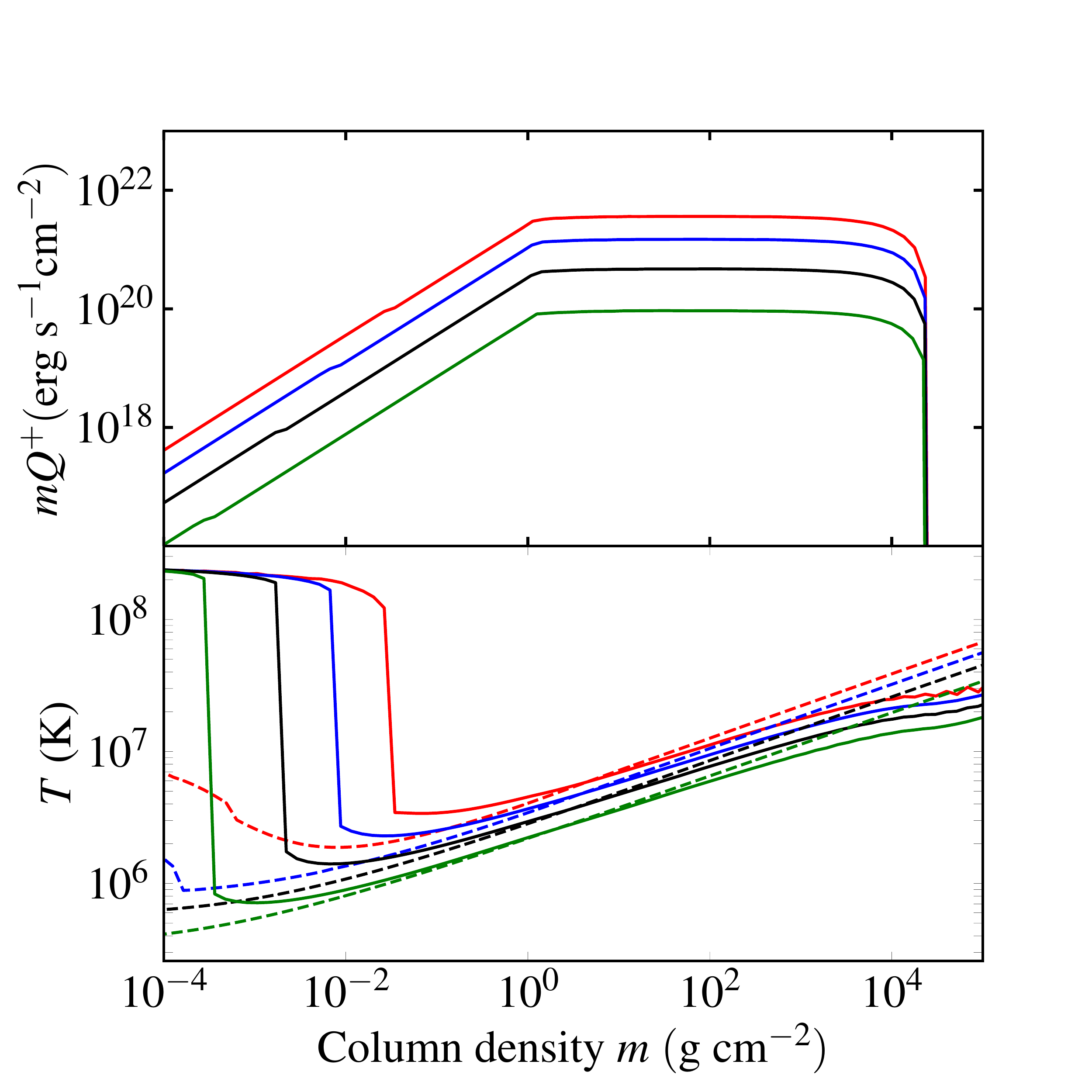} 
\includegraphics[width=8cm]{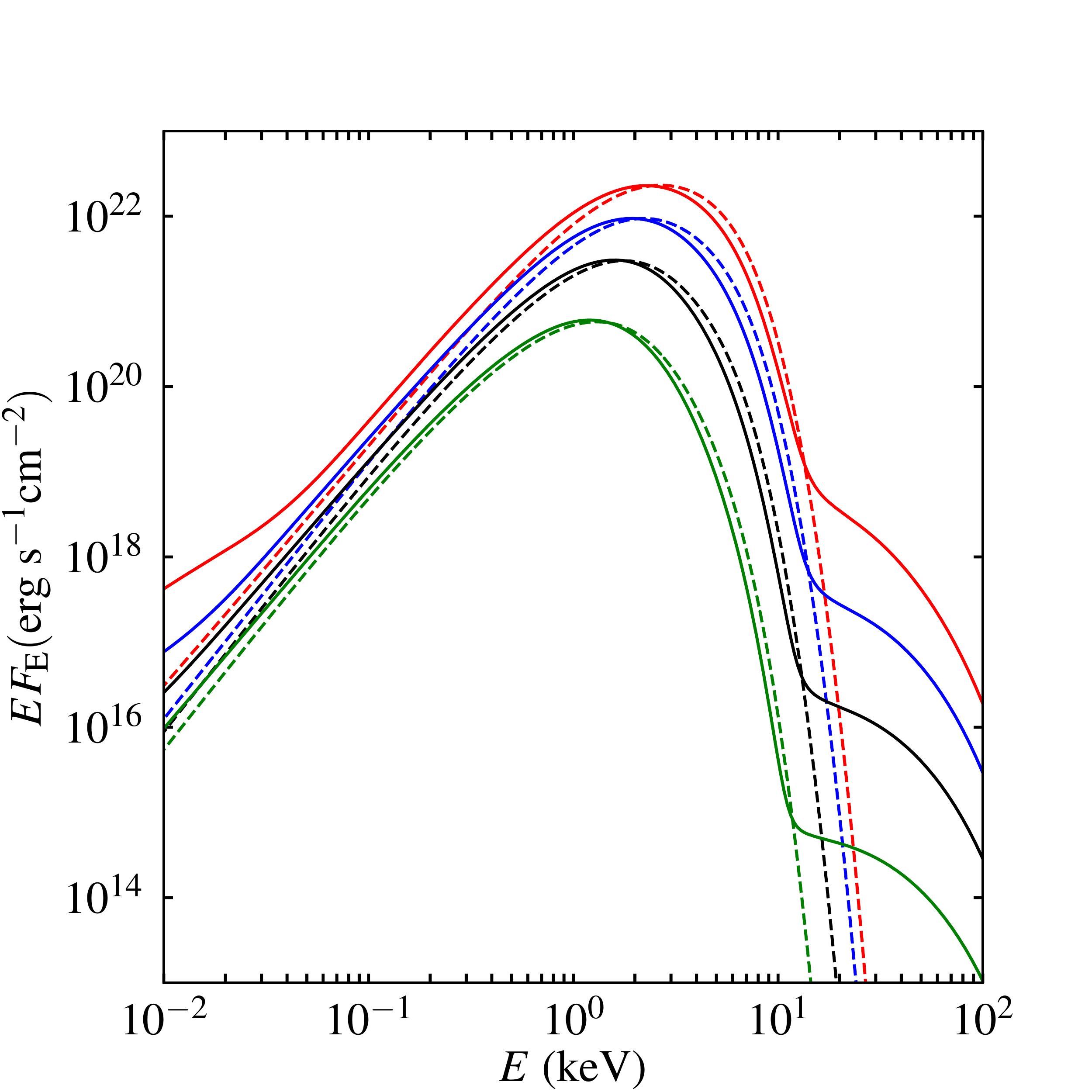} 
\caption{Atmosphere structure (\textit{left}) and the spectrum of escaping radiation (\textit{right}) for a power-law distribution of bombarding particles with $\delta = 2$ and $\gamma_{\mathrm{min}} = 10$ for different NS effective temperatures: $T_{\mathrm{eff,h}} = 2$ (green solid curves), 3 (black), 4 (blue), and 5 MK (red).
Other NS atmosphere parameters are given in Table \ref{table:params}. 
The dashed curves correspond to the temperature structure and escaping spectrum of the atmosphere without external heating (for $T_{\mathrm{eff,i}} = 2$, 3, 4, and 5\,MK). 
}
\label{fig:struc_multiT}
\end{figure*}

Let us now consider a more realistic return-current energy distribution.
More specifically, we used a power-law distribution of bombarding particle energies, expected for realistic magnetosphere return currents \citep{CPS2016}.
We considered different slope $\delta$ of the distribution and studied its influence on the atmosphere structure and the emitted spectra (the fiducial model parameters are shown in Table \ref{table:params}). 
The results for $\delta=1$, 2 and 3 are shown in Fig.~\ref{fig:struc_multiene} (assuming that $\gamma_{\mathrm{min}} = 10$) and compared with corresponding quantities of a non-heated atmosphere. 
In addition, the effect of including the bremsstrahlung energy losses is also shown.
We see that the atmosphere structure and the spectra deviate significantly from those of the non-heated atmosphere if there is contribution of low-energy particles. 

For a steep slope with $\delta = 3$ (corresponding average Lorentz factor $\langle \gamma_{0} \rangle \approx 20$) there are many low-energy particles that heat the surface layers resulting in the 100 keV skin and a significant high-energy photon tail above a few keV. 
On the other hand, the case with $\delta = 1$ correspond to a high average Lorentz factor $\langle \gamma_{0} \rangle \approx 10^{4}$ and the results differ very little from the non-heated case. 
In the case with $\delta=2$ ($\langle \gamma_{0} \rangle \approx 100$), the outer layers of the atmosphere are already largely heated, but the resulting spectra shows major deviations from the non-heated case only at very low and high energies.
All the cases are also well expected based on the comparison of their average Lorentz factors to the mono-energetic results of Sect. \ref{sec:res_monoene}.
In addition, the colour temperature again decreases (as spectral peak shifts) when having lower energy return current particles.
Implications of this (and other deviations from the non-heated model) to NS parameter fitting are discussed in Sect. \ref{sec:discussion_physics}. 

The energy loss due to bremsstrahlung radiation seems not to affect significantly the calculated spectrum and atmosphere structure, except for the temperature profile in the case with $\delta = 1$.
In that case, the inclusion of bremsstrahlung losses makes the highest-energy particles lose their energy significantly faster and at lower depths, which is enough to create a much hotter outer layer.
However, the differences in the spectra are rather small, since the hot skin is not very extended and most of the photons escape from deeper layers.

We also studied how the results depend on the minimum Lorentz factor $\gamma_{\mathrm{min}}$ (assuming $\delta = 2$). 
They are shown in Fig.\,\ref{fig:struc_multigmin} (with $\gamma_{\mathrm{min}} = 10$, $\gamma_{\mathrm{min}} = 50$, and $\gamma_{\mathrm{min}} = 200$, and with no bremsstrahlung losses included).
The lowest energy case (black line) is the same as the intermediate energy case (blue line) of the previous comparison. 
The two other cases correspond to average Lorentz factors $\langle \gamma_{0} \rangle \approx 400$ and $\langle \gamma_{0} \rangle \approx 1400$.

The results show again, that a larger contribution of low-energy particles (smaller $\gamma_{\mathrm{min}}$) leads to hotter upper layers and to a spectrum that deviates more from the non-heated model.
We also note that using the chosen distribution, the choice of the lower limit $\gamma_{\mathrm{min}}$ has a much larger impact on the results than the choice of the upper limit $\gamma_{\mathrm{max}}$.
For example, we tested for $\gamma_{\mathrm{min}}=10$, that $\gamma_{\mathrm{max}}=10^{4}$ would give closely the same structure and spectrum as  $\gamma_{\mathrm{max}}=2 \times 10^{5}$ that was used throughout the paper.
This could also be seen from the results shown in Fig.\,\ref{fig:struc_multiene}, where inclusion of bremsstrahlung energy losses acted in the same way as having significantly smaller $\gamma_{\mathrm{max}}$.

\subsection{Atmospheres with different surface gravities}\label{sec:res_multilgg}

We have also studied the dependency of the results on the surface gravity in the case of distribution of particles with $\delta =2$ and $\gamma_{\mathrm{min}} = 10$.
The results for $\log g = 13.7, 14.0, 14.3$, and 14.6 are shown in Fig.\,\ref{fig:struc_multilgg} (with no bremsstrahlung losses included).
We see that the surface gravity has a much smaller effect on the structure and the spectra of the atmosphere compared to the effects of the energy distribution parameters $\delta$ and $\gamma_{\mathrm{min}}$ of the bombarding particles, discussed in the previous section.
However, the effect is still significant, and similarly as growing $\delta$ or decreasing $\gamma_{\mathrm{min}}$, decreasing the value of $g$ leads to higher amount of hard energy photons.
In this case, the return current energy loss pattern does not change, as seen from the upper left panel of Fig.\,\ref{fig:struc_multilgg}, meaning that the changes in the spectrum and temperature profile are caused by other means (see discussion in Sect.\,\ref{sec:discussion}).
It seems that the lowest surface gravity allows largest inward expansion of the hot skin, resulting in the hardest spectrum.

\subsection{Atmospheres with different effective temperatures}\label{sec:res_multiT}

Additionally, we have considered the dependency of the results on the effective temperature $T_{\mathrm{eff,h}}$ (keeping the ratio $T_{\mathrm{eff,h}}/T_{\mathrm{eff,i}} \approx 3$ same as in the other computations).
The fiducial 5 MK temperature (used in the models of other sections) is relatively high compared to the temperature estimates of recently observed RMPs \citep{GKR_nicer19}.
Therefore, we have calculated the models (again with $\delta =2$ and $\gamma_{\mathrm{min}} = 10$) for $T_{\mathrm{eff,h}} = 2$, 3, 4, and 5 MK. 
The results are shown in Fig.\,\ref{fig:struc_multiT} (solid lines) and compared to the non-heated models of the same effective temperature (dashed lines).
The results show that lower temperature is related to less-pronounced hard tails in the spectra, because for colder atmospheres the temperature inversion happens at lower depths.
This is caused by the cooling rate being more extensively dominated by free-free radiation instead of Compton scattering (see discussion in Sect.\,\ref{sec:discussion}.).
However, the increased emission due to heated upper layers can still be detected at high energies even for $T=2$ MK.

\begin{figure*}
\centering
\includegraphics[width=7.5cm]{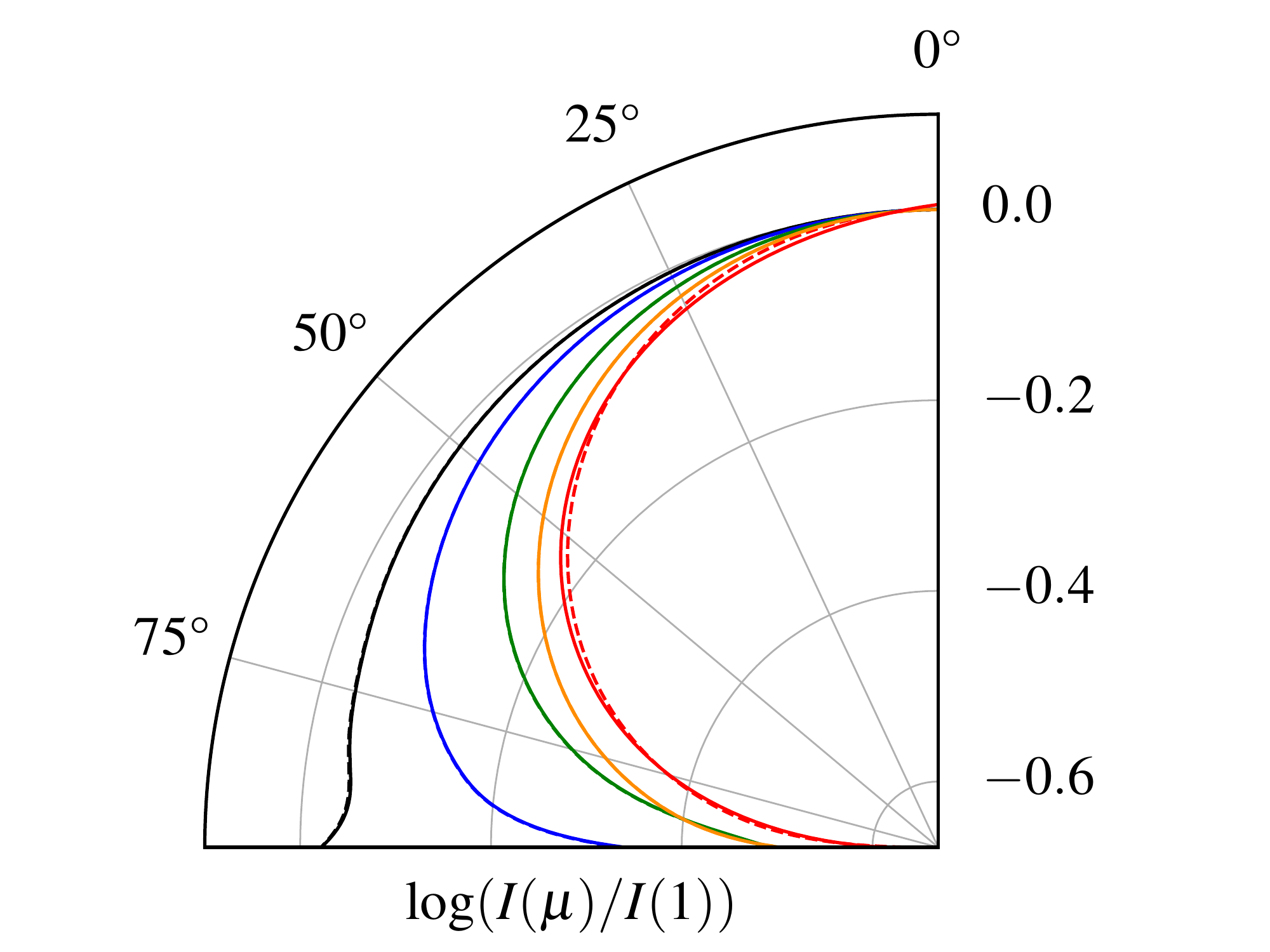} 
\includegraphics[width=7.5cm]{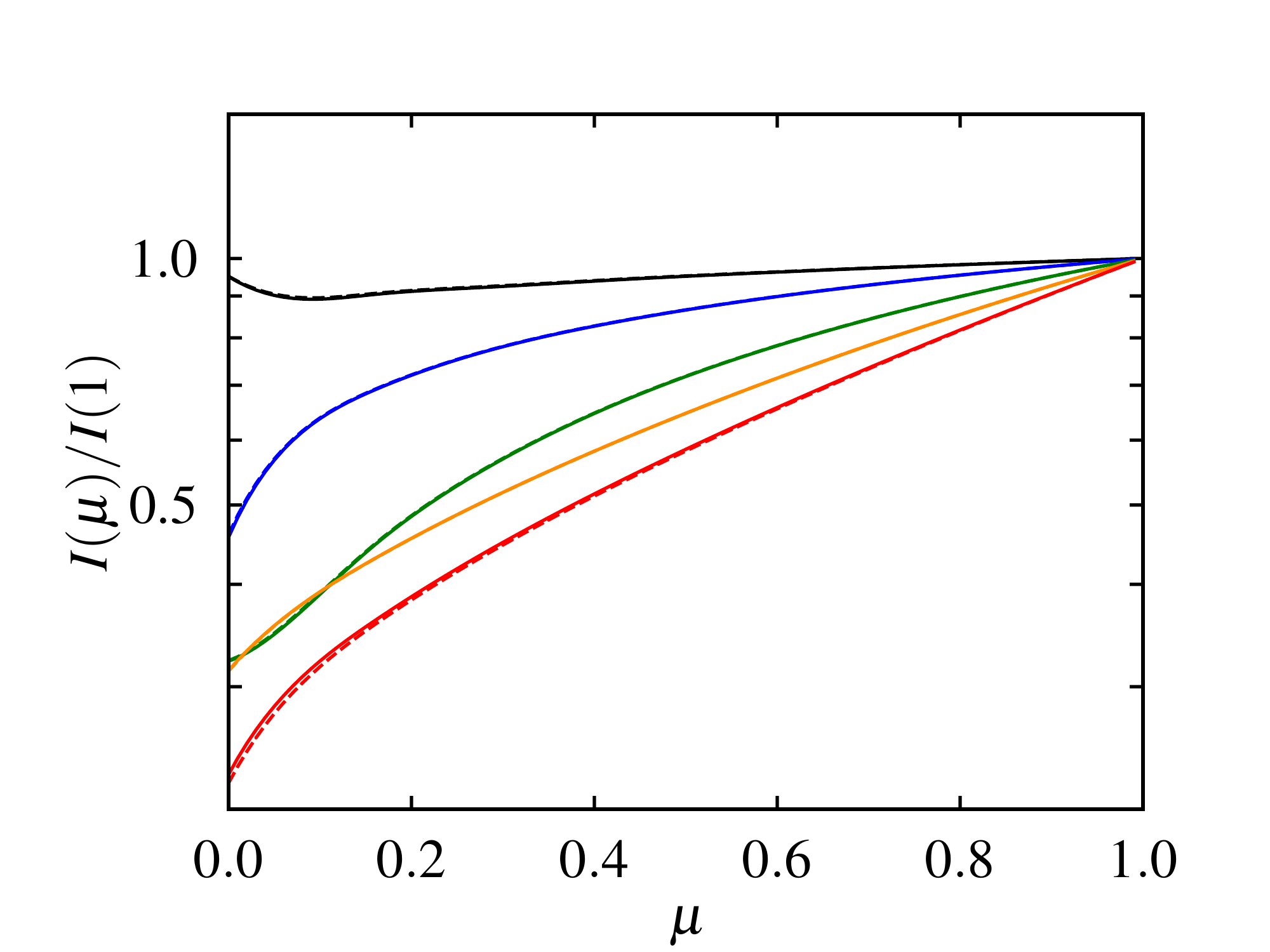} 
\includegraphics[width=7.5cm]{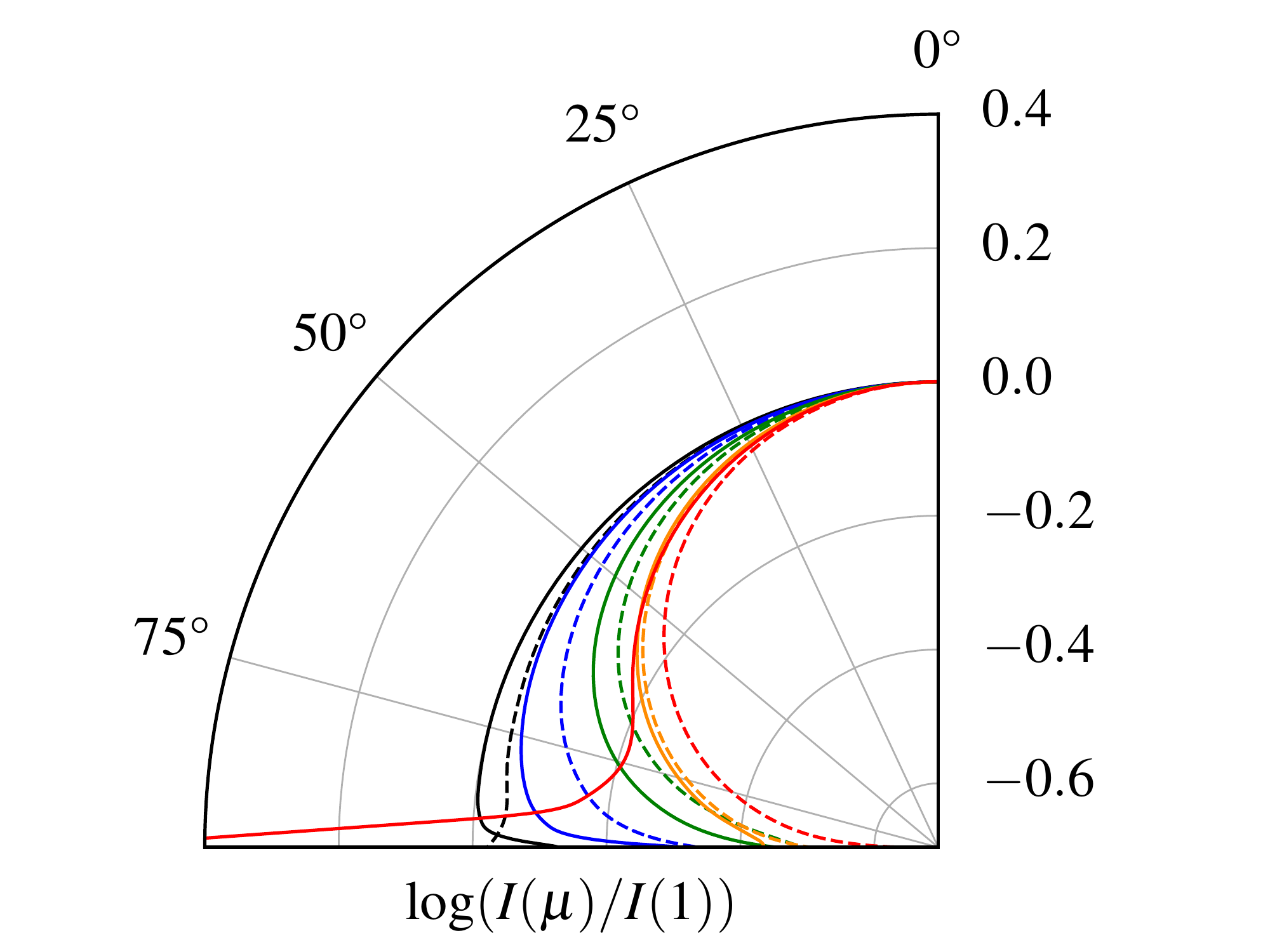} 
\includegraphics[width=7.5cm]{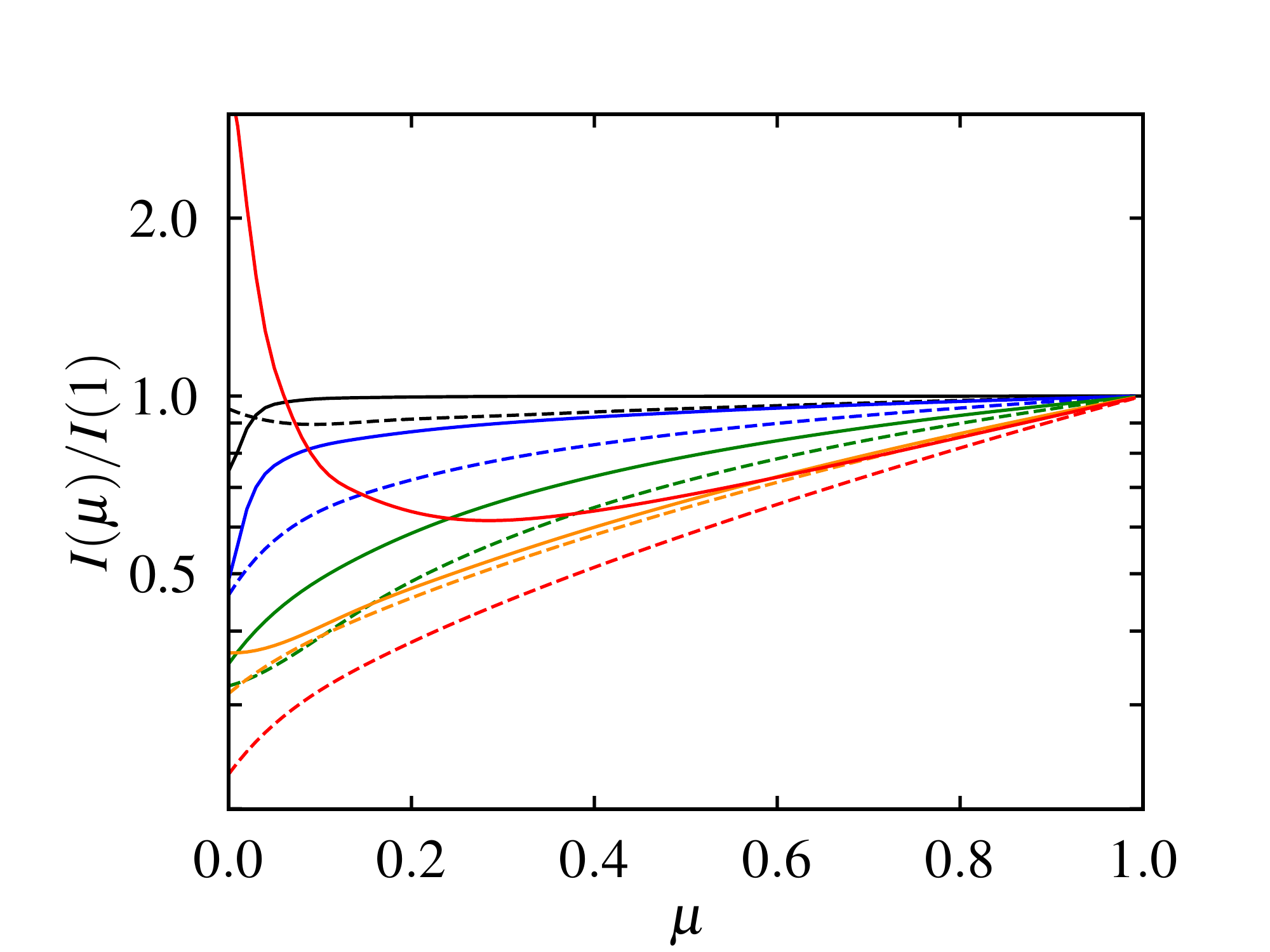} 
\includegraphics[width=7.5cm]{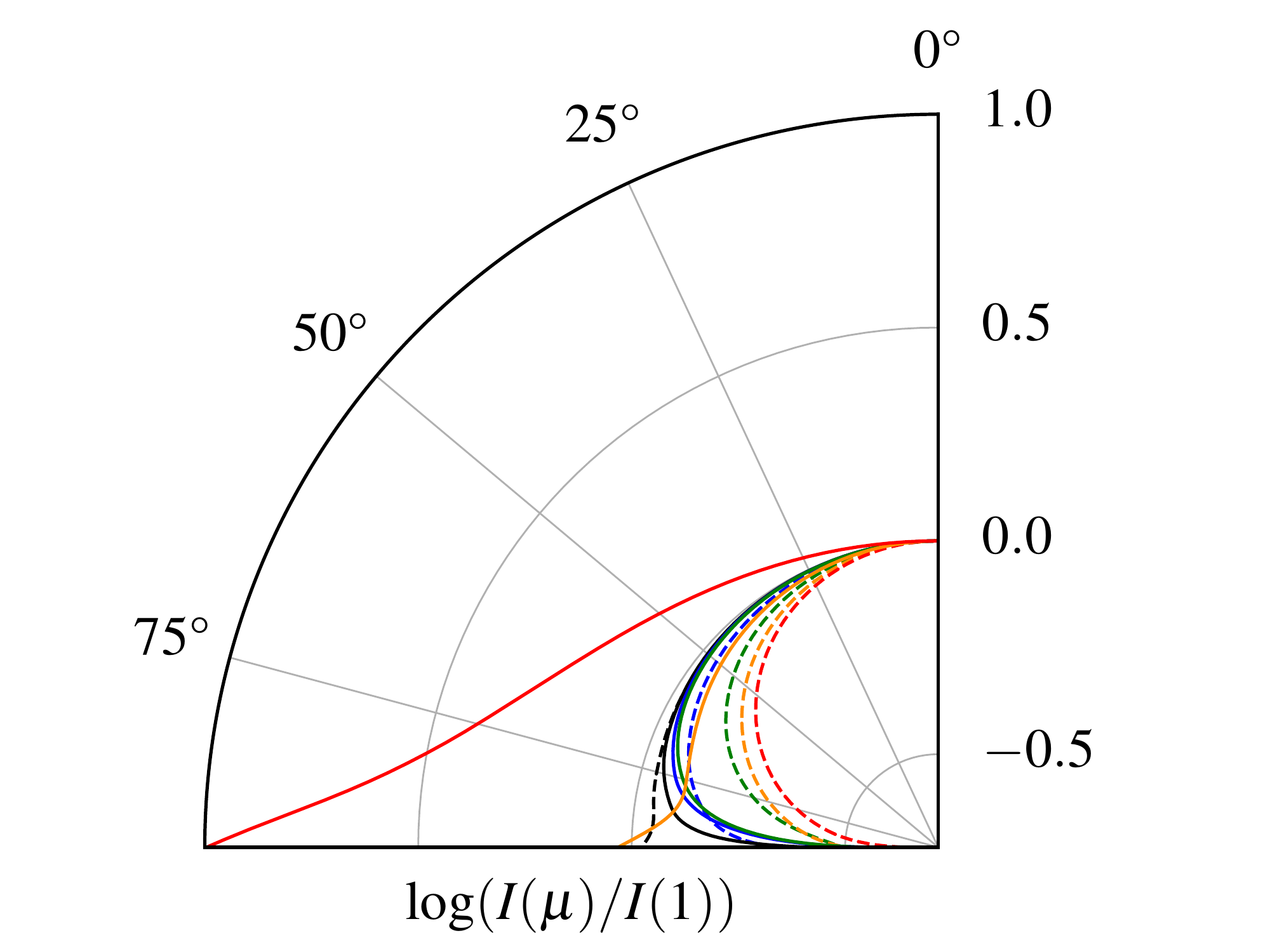} 
\includegraphics[width=7.5cm]{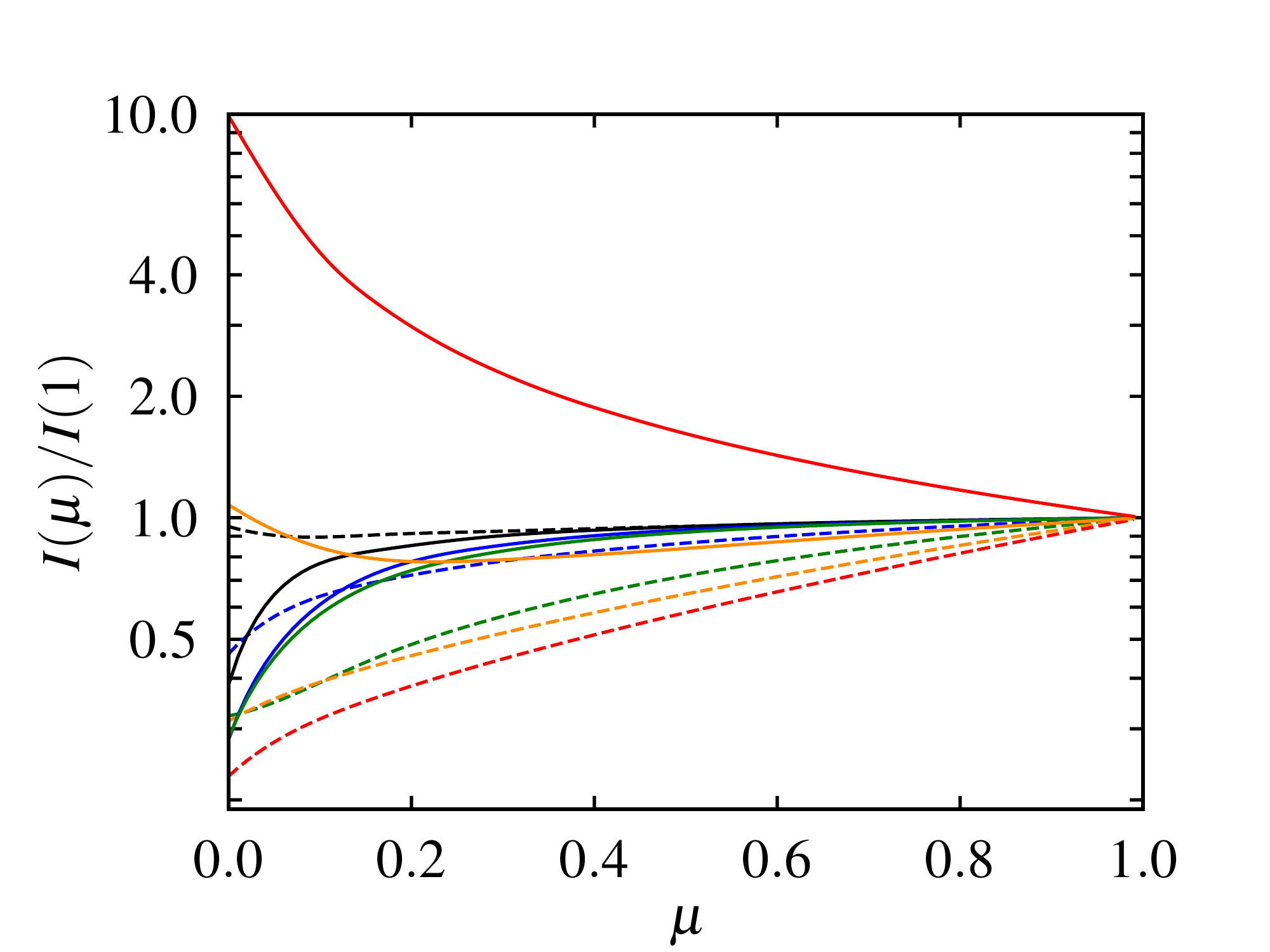} 
\caption{
Emission pattern of the specific intensity in polar (\textit{left}) and Cartesian (\textit{right}) coordinates for the NS atmosphere models presented in Fig.\,\ref{fig:struc_multiene}. 
The dashed curves show the patterns for non-heated atmosphere model, and the solid curves are for a power-law distribution of return current particles. 
\textit{Top, middle and bottom panels} correspond to $\delta = 1$, 2 and 3, respectively. 
The black, blue, green, orange, and red colours correspond to intensities at 0.1, 0.3, 1, 3, and 10 keV, respectively. 
}
\label{fig:angdep_multiene_d1}
\end{figure*}

\subsection{Beaming patterns}

Let us now discuss the angular distribution of the emergent radiation.
The emission pattern of radiation escaping from the atmosphere of $\log g = 14.3$ and $T_{\mathrm{eff,h}} = 5$ MK for a power-law distribution of in-falling particles with $\gamma_{\mathrm{min}} = 10$ and three slopes $\delta = 1$, 2 and 3 is shown in both in polar (left) and Cartesian (right) coordinates in  Fig.\,\ref{fig:angdep_multiene_d1}.
The specific intensities at different cosines of zenith angles (denoted as $\mu$) were obtained directly using the radiative transfer equation, that was solved finally in eleven angles (instead of three angles used during the major part of temperature iterations). 
The intensities for the intermediate angles were obtained using a third order spline interpolation from the calculated points. 

We see that in the case of low-energy particle heating (i.e. $\delta=3$) the beaming of the radiation at low and high energies is clearly different, unlike in the case of non-heated atmosphere.
The energy, where the limb darkening at low energies changes to the limb brightening at higher energies, depends on the return-current electron energy distribution.
In case of $\delta = 1$, the angular distribution of radiation resembles very closely the distribution of a non-heated atmosphere, and no limb brightening is seen. 
When $\delta = 2$, we observe a rapid transition from limb darkening to limb brightening at the highest zenith angles and highest energies (10 keV). 
Even in the thermal part of the spectrum, we see large deviations of the angular distribution from that predicted by standard non-heated model. 
With energy distribution having even higher small-$\gamma$ contribution (i.e. $\delta=3$), the energies above 3\,keV show even stronger brightening, for all angles, towards the surface tangent.
The deviations between heated and non-heated models become slightly larger also at the smaller energies, when a significant fraction of the energy of the beam is dissipated high up in the atmosphere ($\delta \geq 2$).

\begin{figure}
\centering
\includegraphics[width=8cm]{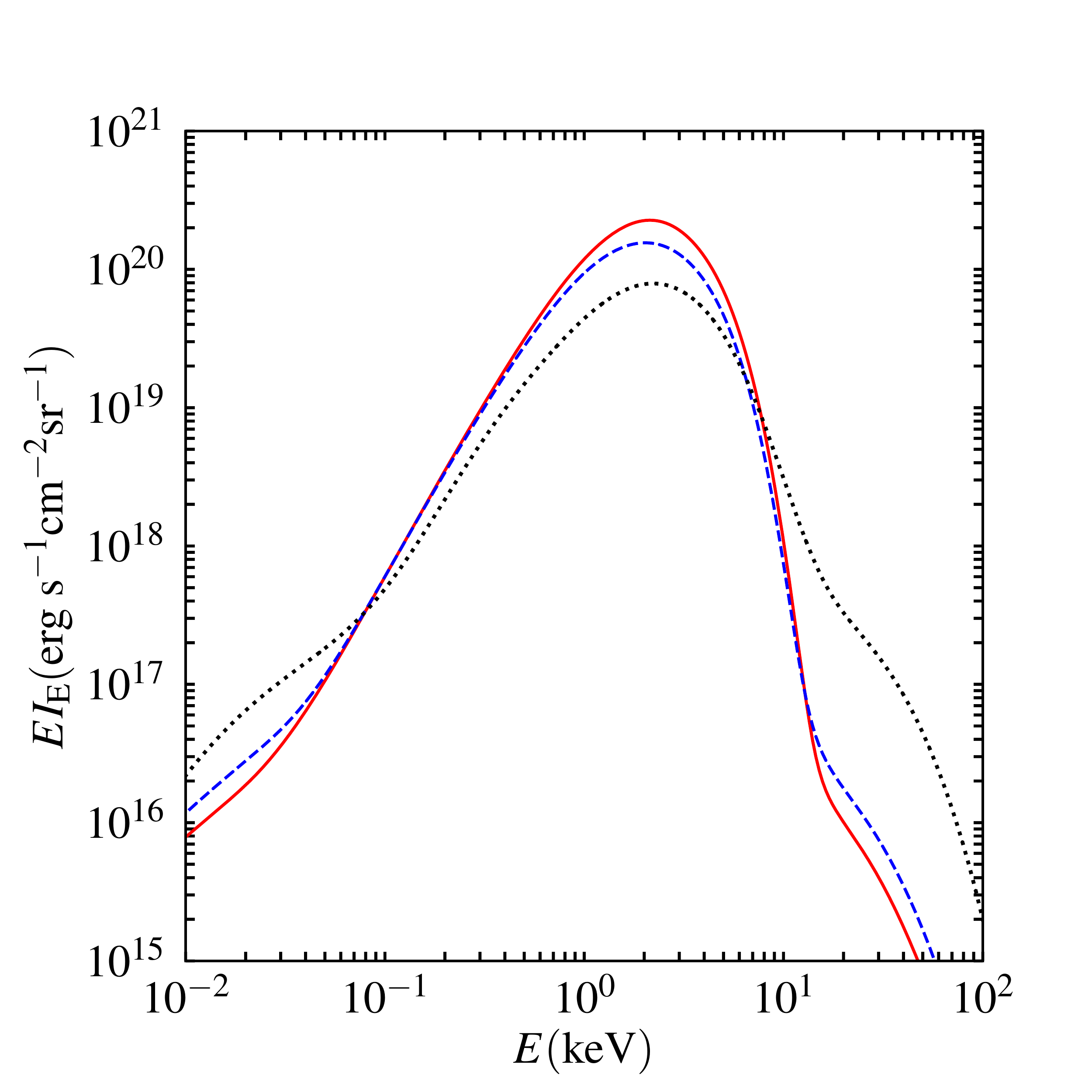} 
\caption{Emergent specific intensity spectra for power-law distribution of the particle beam for the NS atmosphere model shown in Fig.\,\ref{fig:struc_multiene} in case of $\delta=2$ and $\gamma_{\mathrm{min}}=10$.
The results are shown for three emission angles (out of the eleven that were computed): $\mu = 0.99$ (red solid curve), $\mu = 0.50$ (blue dashed), and $\mu = 0.01$ (black dotted).
}
\label{fig:angspec_multiene}
\end{figure}

The emergent spectra for different zenith angles are shown in Fig.\,\ref{fig:angspec_multiene}.
From this we see that the high-energy tail of the spectrum is larger for high zenith angles, as expected.
On the other hand, the peak of the spectrum is higher for small angles. At very low energies again the radiation intensity becomes slightly larger at high zenith angles.

\begin{figure}
\centering
\includegraphics[width=8cm]{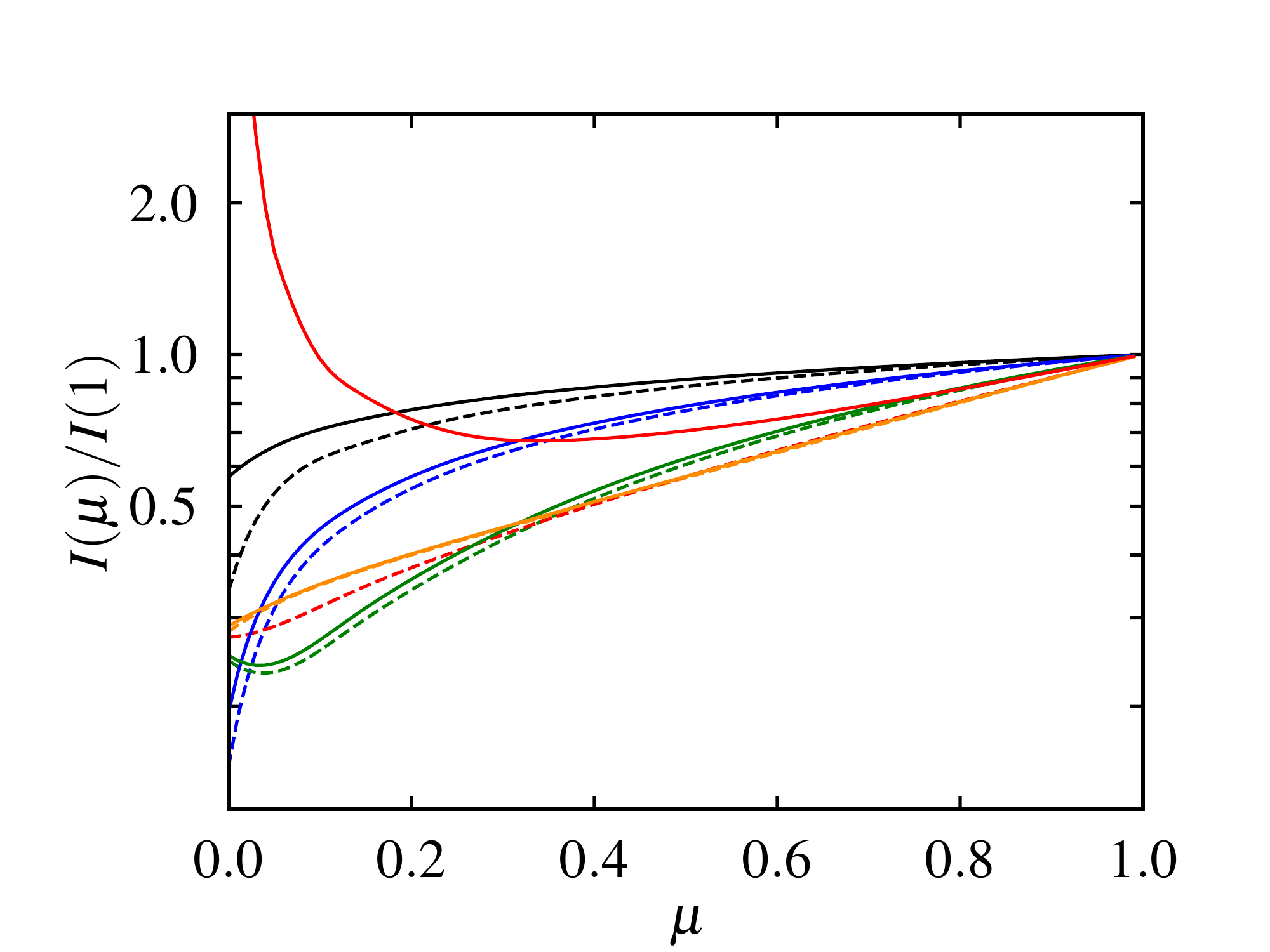} 
\caption{
Emission pattern of the specific intensity for the NS atmosphere models in a low temperature case with $T_{\mathrm{eff}} = 2$\,MK (with $\delta=2$) presented in Fig.\,\ref{fig:struc_multiT}. 
The dashed curves show the patterns for non-heated atmosphere model, and the solid curves are for a power-law distribution of return current particles. 
The black, blue, green, orange, and red colours correspond to intensities at 0.1, 0.3, 1, 3, and 10 keV, respectively.
}
\label{fig:angdep_T2}
\end{figure}

We also found similar discrepancies between the beaming patterns of heated and non-heated models in case of a colder atmosphere ($T_{\mathrm{eff,h}}$ = 2 MK or without heating $T_{\mathrm{eff,i}}$ = 2 MK). 
The results are shown in Fig.\,\ref{fig:angdep_T2}.
We see again rapid transition from limb darkening to limb brightening at 10 keV, although the energies should not be directly compared to those computed with different temperature (since the thermal peak is shifted).
Nevertheless, the differences in the angular distributions remain at 10--30\% level even at smaller energies.
This demonstrates that the effects of return-current-heating effects should be relevant also for RMPs with relatively low temperatures studied recently \citep{MLD_nicer19,RWB_nicer19}.

The differences between the beaming patterns can also be coarsely characterised by inspecting the values of intensity emitted towards tangential direction $\mu=0$.
A summary of all the computed models (excluding those where bremsstrahlung losses and mono-energetic particles were considered) and the values obtained for the normalised intensity at $\mu = 0$ (defined as $a = I(\mu=0)/I(\mu=1)$) for four different energies (0.3, 1, 3, and 10 times the colour temperature $kT_{\mathrm{c}}$ of each model) are shown in Table\,\ref{table:all_models}.
The colour temperatures (and the colour correction factors $f_{\mathrm{c}}$) were obtained by fitting a diluted blackbody function to the emergent spectrum.
The fitting procedure was similar to the first method described in \citet{SPW11}, although the energy band was adjusted to $(0.1-6)\times(1+z)$ keV (for RMPs detected by NICER), where $z$ is the gravitational redshift, obtained from $\log g$ by assuming a NS mass of $1.4~\msun$.

We can see that the angular dependency of the intensity is different from non-heated models also when making the comparison at the same energies relative to the spectral peak (see Table\,\ref{table:all_models}).
The location of the peak varies between the models, so that more over-heated upper layers, or higher effective temperature, produce smaller $f_{\mathrm{c}}$.
From Table\,\ref{table:all_models} we also see that the beaming at large angles is most insensitive (although not entirely) to the used model at close to the peak of the spectrum ($a_{3}$ at $3kT_{\mathrm{c}}$). 
At these energies the angular dependency can be approximated by a linear function as $I(\mu)/I(1) = a + (1-a)\mu$.
For energies above and below the spectral maximum, the deviations in the beaming become larger, and the approximately linear dependency breaks down. 

\begin{table*}
\begin{center}
 \begin{minipage}{170mm}
  \caption{Colour correction factors and beaming parameters for the computed atmosphere models.   }\label{table:all_models}
\centering
  \begin{tabular}[c]{c c c c c c c c c c c} 
    \hline\hline
      $T_{\mathrm{eff}}$ (MK) & $\log g$  & $\gamma_{\mathrm{min}}$ & $\delta$ & $\langle \gamma_{0} \rangle$ & $n_{\mathrm{in}}/n_{\mathrm{GJ}}$ & $f_{\mathrm{c}}$ & $a_{1}$ &  $a_{2}$ & $a_{3}$ & $a_{4}$ \\ \hline   
\multicolumn{11}{c}{Heated models} \\  
$5$ & $13.7$ & $10$ & $2$ & $100$ & $2 \times 10^{5}$ & $1.15$ & $0.44$ & $0.38$ & $0.34$ & $0.59$ \\
$5$ & $14.0$ & $10$ & $2$ & $100$ & $2 \times 10^{5}$ & $1.20$ & $0.58$ & $0.42$ & $0.34$ & $0.55$ \\
$5$ & $14.3$ & $10$ & $2$ & $100$ & $2 \times 10^{5}$ & $1.26$ & $0.59$ & $0.42$ & $0.34$ & $0.54$ \\
$5$ & $14.6$ & $10$ & $2$ & $100$ & $2 \times 10^{5}$ & $1.33$ & $0.84$ & $0.45$ & $0.33$ & $0.43$ \\
$5$ & $14.3$ & $10$ & $1$ & $10^{4}$ & $2 \times 10^{3}$ & $1.51$ & $0.64$ & $0.32$ & $0.32$ & $0.29$ \\
$5$ & $14.3$ & $10$ & $3$ & $20$ & $10^{6}$ & $0.93$ & $0.38$ & $0.28$ & $0.32$ & $2.40$ \\
$5$ & $14.3$ & $50$ & $2$ & $400$ & $5 \times 10^{4}$ & $1.42$ & $0.80$ & $0.36$ & $0.32$ & $0.33$ \\
$5$ & $14.3$ & $200$ & $2$ & $1400$ & $10^{4}$ & $1.48$ & $0.62$ & $0.33$ & $0.32$ & $0.31$ \\
$4$ & $14.3$ & $10$ & $2$ & $100$ & $9 \times 10^{4}$ & $1.36$ & $0.87$ & $0.44$ & $0.33$ & $0.41$ \\
$3$ & $14.3$ & $10$ & $2$ & $100$ & $3 \times 10^{4}$ & $1.49$ & $0.86$ & $0.35$ & $0.30$ & $0.34$ \\
$2$ & $14.3$ & $10$ & $2$ & $100$ & $5 \times 10^{3}$ & $1.67$ & $0.64$ & $0.21$ & $0.23$ & $0.29$ \\
\multicolumn{11}{c}{Non-heated models} \\
$5$ & $14.3$ & $-$ & $-$ & $-$ & $-$ & $1.52$ & $0.64$ & $0.32$ & $0.32$ & $0.31$ \\
$4$ & $14.3$ & $-$ & $-$ & $-$ & $-$ & $1.59$ & $0.43$ & $0.27$ & $0.31$ & $0.31$ \\
$3$ & $14.3$ & $-$ & $-$ & $-$ & $-$ & $1.67$ & $0.41$ & $0.21$ & $0.29$ & $0.31$ \\
$2$ & $14.3$ & $-$ & $-$ & $-$ & $-$ & $1.80$ & $0.37$ & $0.16$ & $0.23$ & $0.28$ \\
\hline 
  \end{tabular}
\end{minipage}
\end{center}
\tablefoot{The model parameters are: the effective temperature $T_{\mathrm{eff}}$, surface gravity $\log g$, minimum Lorentz factor of the incoming particles $\gamma_{\mathrm{min}}$, and the slope of the return-current particle energy distribution $\delta$.
Also, shown are the average Lorentz factor $\langle \gamma_{0} \rangle$, number density of penetrating particles in terms of Goldreich-Julian number
density $n_{\mathrm{in}} / n_{\mathrm{GJ}}$, colour correction factor $f_{\mathrm{c}}$  (defined as $f_{\mathrm{c}} = T_{\mathrm{c}}/T_{\mathrm{\mathrm{eff}}}$, where $T_{\mathrm{c}}$ is the fitted colour temperature), and beaming parameters $a_{1}$, $a_{2}$, $a_{3}$, $a_{4}$ (where $a_{\mathrm{i}} = I_{\mathrm{i}}(\mu=0)/I_{\mathrm{i}}(\mu=1)$) for photons emitted at energies 0.3, 1, 3, and 10 $kT_{\mathrm{c}}$ correspondingly. 
}
\end{table*}

\section{Discussion}\label{sec:discussion}

\subsection{Difference from deep-heating models and implications for NICER results}
\label{sec:discussion_physics}

The results presented in Sect.\,\ref{sec:results} demonstrate that the spectra and beaming patterns for the return-current-heated atmosphere models can significantly differ from those assuming that heat is released at the bottom of the atmosphere (non-heated models).
As we have shown, the results depend on the unknown energy distribution of the return-current particles.
The difference is very large for the smallest energies of the bombarding particles, but insignificant for the highest energies that we have considered.
Even when the spectral differences are small (e.g. for the cases with $\gamma_{0} = 500$, or $\delta=1$, or $\gamma_{\mathrm{min}} = 200$), the two-layer structure with an overheated upper layers can still be produced.
The heated part of the atmosphere consists of a hot optically thin skin and a cooler optically thick inner part. 
The increased temperature, due to the return current, forces the outer layers to expand, as the pressure does not change, and creates a discontinuity in the particle concentration at the depth where the temperature inversion occurs. 
This allows for a possibility to represent the RMP atmosphere with a simplified two-layer model consisting of a hot Comptonizing slab above a usual deep-heated atmosphere.

We note that the depth of the temperature inversion is likely related to the change in the dominant opacity mechanism.
The cooling rates due to free-free radiation and due to Compton scattering are almost equal always at the point of the temperature inversion.
In case of a higher surface gravity, the free-free opacity dominates at larger range of depths in the atmosphere (because of higher gas pressure and electron number densities for the same temperature) and therefore the temperature inversion occurs also at a lower depth and the spectrum is closer to the blackbody (as seen in Fig.\,\ref{fig:struc_multilgg}).
In case of a fixed surface gravity, the amount of energy deposited in the upper layers determines the inversion depth, so that the inversion occurs at higher depth and outer layers become hotter when the amount of the deposited energy in the upper layers is increasing. 
Nevertheless, most of the observed radiation still escapes from the layers deeper than the inversion depth in all of our models.
For models with less energetic bombarding particles, the layers producing thermal radiation are less efficiently heated, and that can explain why the peak of the spectrum is shifted towards the lower energies (see Figs. \ref{fig:struc_gall}--\ref{fig:struc_multigmin}).  
As a consequence, this leads to a substantial bias in the estimate of the atmosphere effective temperature (and also in the size of the emitting spot) when constraining NS parameters using phase-resolved spectra, if incorrect assumptions about the atmosphere model are made.  

As we have shown above, the radiation spectrum and the angular distribution of escaping radiation may deviate significantly from those predicted by standard hydrogen atmosphere models without external heating. 
What is important is that the angular distribution even of the thermal part of the spectrum (see blue and green curves in the middle panel of Fig.\,\ref{fig:angdep_multiene_d1} and in Fig.\,\ref{fig:angdep_T2}), when the high-energy tail is completely negligible (see Fig.\,\ref{fig:struc_multiT}), shows 10--50\% difference from the corresponding angular distributions of non-heated models.
Obviously, a different emission pattern of the hotspots on the surface of RMPs would produce a significantly different pulse profile \citep[see e.g.][]{PB06}, with characteristic difference at the tens of per cent level. 
This is orders of magnitude larger than the accuracy of 0.1\% the NICER team aims at in their pulse profile modelling \citep{BLM_nicer19}.
Because in both NICER papers on NS parameters constraints from PSR J0030+0451 \citep{MLD_nicer19,RWB_nicer19} the spectral shape and the angular distribution were taken from standard non-heated models, the obtained constraints on the NS radius and the structure of magnetosphere may be heavily biased. 
Detailed analysis and improved constraints on the NS parameters and magnetospheric geometry obtained from the pulse profile modelling of RMPs using 
return-current-heated atmosphere models will be considered in a forthcoming publication.

Conversely, investigation of the thermal spectra of RMPs opens a possibility to study the energy distribution of the relativistic particles by using the atmosphere model to constrain the particle distribution parameters.
This would allow us to also indirectly probe the exact physics of pulsar magnetospheres via the return-current structure.
Such studies can provide an interesting new observational method for actually validating, for example, the different proposed pulsar magnetosphere gap models or pair-production sites in the magnetosphere since both of these physical details alter the incoming return current properties \citep[see e.g.][] {Chen2014,Cerutti2017,Beskin2018,Philippov2018}.

\subsection{Comparison with previous works}
\label{sec:discussion_comparison}

Compared to the results of \citet{BPO19}, who also studied the return-current-heated atmospheres, our calculations show similar features in the spectra and beaming patterns, although having a significantly higher temperature of the outer atmosphere layers, using the same parameters for the energy distribution of the return current.
Thus, also our spectrum deviates more from a blackbody, typical for a non-heated atmosphere.
In our case, spectra are harder and the resulting beaming pattern slightly more limb brightened at high energies for low return-current electron energies.

The main difference of our atmosphere model compared to that of \citet{BPO19} is in the modelling of the radiative transfer.
We account exactly for the effects of electron scattering and perform calculations at a grid of photon energies, while  \citet{BPO19} consider grey atmosphere. 
This is likely the reason why the atmosphere structure and the resulting spectra are significantly different. 
As a part of accurate radiation transfer modelling, we also use energy-dependent opacities instead of averaged ones. 
We checked that in the overheated layers the Planck-weighted opacity differs significantly from the opacity that is weighted using correct intensities. 

Qualitatively, our results are similar to those of \citet{SPW18}, who instead of RMPs, studied the accretion-heated atmospheres of NS. 
The angular distribution of the emergent radiation is comparable to ours, as we see the limb darkening at lowest energies switching to the limb brightening at higher energies.
In addition, they have the pronounced two-layer structure of the atmosphere, and emergent spectra showing a similar excess at lower photon energies in comparison with the blackbody.
This is a consequence of the free-free opacity being high at low energies causing the observed low energy photons to be produced in the overheated upper layers. 
This ``reverse photosphere effect'' was already discussed by \citet{DDS2001}, who also obtained similar structures and spectra using a pure hydrogen atmosphere heated by accreted protons. 
It could also be possible to detect the low energy excess as an increased emission at optical wavelengths. 
However, the observed optical excess, usually associated with highly magnetised neutron stars \citep{vkk07} may be explained by magnetic atmosphere models instead \citep{Ho2007,GCZT19}.

\subsection{Connection to pulsar physics}

In pulsar physics, instead of the effective temperature $T_{\mathrm{eff,h}}$, the parameter of interest is the number  density of precipitating particles in terms of the Goldreich-Julian number density defined as \citep{GJ69} 
\be 
n_{\mathrm{GJ}} \approx  \frac{B}{ePc},
\ee 
where $B$ is the magnetic field strength, $P$ is the period of the pulsar, and $e$ is the elemental charge.
It defines a characteristic density in the pulsar magnetosphere that is needed to screen the longitudinal electric field near the neutron star surface \citep{Beskin2018}.
The pulsar magnetosphere has a total number density $n_{\mathrm{tot}} = \mathcal{M} n_{\mathrm{GJ}}$, where $\mathcal{M}$ is the pair multiplicity parameter. 

The number density of the inward penetrating particles suggested by our atmosphere models (see Eq. \eqref{eq:number_flux}), on the other hand, is given as 
\be \label{eq:number_in}
n_{\mathrm{in}} \approx \frac{\dot{N}_{\mathrm{e}}}{c} = \frac{ \sigma_{\mathrm{SB}}T_{\mathrm{eff,h}}^{4}}{(\langle \gamma_{0} \rangle -1)m_{\mathrm{e}}c^{3}} 
=\zeta \mathcal{M} n_{\mathrm{GJ}},
\ee
where $\zeta = n_\mathrm{in}/n_\mathrm{tot}$ is the ratio of in-going to the total number of pairs.
Both $\zeta$ and $\mathcal{M}$ are currently unknown and depend on the exact details of pulsar magnetosphere structure and pair cascade physics.

Multiplicities corresponding to our models are shown in Table \ref{table:all_models}, when assuming typical values for millisecond pulsars ($B=10^{9}$ G and $P=1$ ms). 
For simplicity, we have also taken $\zeta = 1$.
The multiplicity depends on our model parameters $T_{\mathrm{eff,h}}$, $\delta$ and $\gamma_{\mathrm{min}}$ (the two latter through $\langle \gamma_{0} \rangle$) as seen from Eq. \eqref{eq:number_in}.
The presented values range from $\mathcal{M} \sim 10^{3}$ to $10^{6}$ which are reasonable given the theoretical uncertainties in pulsar magnetosphere physics \citep{TA13}. 

We note that the already large uncertainties of $\zeta$ and $M$ are increased when considering millisecond pulsars.
The details of the pair-production mechanism in millisecond pulsars must differ from slower rotating pulsars that can enable efficient pair production via a strong magnetic field;
in millisecond pulsars the process could be seeded, for example, by high-energy inverse Compton photons instead.
However, to have at least a modest view on the return current physics and particle distribution, we briefly consider next our models in the context of recent developments in simulations of rotation-powered pulsars.

One-dimensional particle-in-cell (PIC) simulations of the polar cap pair cascades, created near the magnetic poles, have shown that a large fraction of the energy may be carried by particles having very high Lorentz factor of $\gamma \sim 10^7$ \citep{T10, TA13}. 
However, in these studies also a significant population of low-energy particles flow towards the NS surface heating the upper layers of the atmosphere.
Global multi-dimensional PIC simulations of pulsar magnetospheres has been performed as well, but using scaled energies of the return current particles \citep{Chen2014,Cerutti2017,Philippov2018}.
Therefore, the typical Lorentz factor of secondary pairs cannot be directly inferred.
However, the particle energy distribution depends on the details of the model and the assumed type of the pulsar.  
These uncertainties translate to a large uncertainty in the $\zeta$ parameter, which could be, for example, around 0.1 for strong-field pulsars \citep{TA13}.

One-dimensional PIC simulations have also indicated that pair creation must be time-dependent and exhibit a quasi-periodic behaviour in accelerators where the pair formation is not suppressed \citep{BT07,Belo08,TA13}. 
The non-stationary nature of the process means that the physically relevant parameters to consider are, in reality, time-averaged quantities $\bar{\zeta}$ and $\bar{M}$, if the characteristic relaxation time scale of the atmosphere is longer than time scale of the fluctuations.
The latter may be from fractions of microseconds to tens of microseconds depending on the altitude where the cascade starts \citep{TA13}.
The thermal relaxation time scale of the atmosphere can also be about the same order of magnitude, depending however on the model.
Therefore, a demand for time-dependent atmospheres models (which are not considered here) cannot be ruled out.
However, we note that if the amplitude of the fluctuations is small enough, as it may be in multi-dimensional simulations, the atmosphere is not affected.

\subsection{Caveats}
\label{sec:discussion_caveats}

There are a few uncertainties in our calculated model atmospheres.
One is our assumption that all the dissipated energy (from the return current particles) translates to the outward directed flux at every depth point, as can be seen in our definition of the theoretical flux (the nominator in Eq. \eqref{eq:rel_flux_error}).
The same assumption is implicitly used also in our energy balance correction, as the sign of $Q^{+}$ has been chosen to correspond to a negative flux derivative implying an outward flux.
The assumption might not be physical, and inward flux could occur, for instance, just below the stopping depth of the penetrating particles, if the initial heat of the NS is significantly smaller than heat due to the stopped particles.
However, considering the internal heating of the NS, caused by the return current, is not in the scope of this paper.
Also, if all the incoming energy is lost at optical depths smaller than 1, as seems to be in our models, most of the energy is expected to be radiated outwards anyway.

In our analysis we neglected the effects of thermal conduction, affecting the heating and cooling rate of the matter. 
However, based on the results of \citet{SPW18} this effect is expected to be rather small.
Also, the electron-positron pair production effects were neglected.
For some of the model parameters, where we obtain outer layers as hot as $T = 10^{9}$ K, we could have an outflow of pairs \citep{ztt98}.
Detailed study of this effect is left for a future work.
Furthermore, as we mentioned above, the bombarding particles are actually positrons, and therefore they are annihilated with the background electrons once their energy is low enough, producing an annihilation line at 511 keV. 
Half of those photons may reach the observer directly and another half are Compton back-scattered, making an extended tail containing more than $1/\gamma_0$ fraction of the total energy.

Another simplification in our model is related to the bremsstrahlung radiation as a source of heating. 
We assume that all the bremsstrahlung radiation of the stopping particles is converted to heat of the surrounding gas at the same optical depth. 
This assumption should be valid in the deepest layers (where generated high-energy photons rapidly lose their energy due to Compton scattering) but some deviations in the optically thin upper layers could be expected (photons would rather heat the layers below them).
However, as seen from those results where bremsstrahlung effects were included, the resulting spectra and temperature structure do not have major changes because of this energy loss mechanism.
In addition, the bremsstrahlung radiation should still be taken into account as an additional source of photons in the radiation transfer equation. 
The inward moving high-energy bremsstrahlung photons produced at relatively low depths are Compton back-scattered causing a bump at energies above the thermal peak. 
This effect is most important when bremsstrahlung energy losses are highest and the spectrum would otherwise look thermal, like in non-heated models.

A major uncertainty lies in the energy spectrum of the magnetospheric particles, especially in the low-$\gamma$ regime as discussed also by \citet{BPO19}. 
The spectra for the particle beam could deviate from the power-law. 
There could be, for example, different contribution of low-$\gamma$, either because of different shape of the distribution, or because of using different lower $\gamma_{\mathrm{min}}$ and the upper limits  $\gamma_{\mathrm{max}}$ of the distribution.
However, this uncertainty was considered in this study, as we examined the results using different values for the distribution parameters.

\section{Conclusions}
\label{sec:conclusions}

We have presented a model for return-current-heated atmospheres of RMPs, using the full radiative transfer calculation and accounting exactly for Compton scattering.
We have assumed that the magnetospheric return current that heats the polar caps of NSs consists of pair plasma and considered different energy distributions.
We compared various energy loss mechanisms of the return current, concluding that the effect of the bremsstrahlung energy loss is small.
Finally, we computed the temperature structures, emergent spectra and angular distribution of radiation for different model parameters.

We found that the usual deep-heating approximation deviates significantly from the return-current-heated model if the bombarding particles have a significant contribution at Lorentz factors smaller than about 100. 
The models including high contribution of low-energy particles resulted in a very hot skin, reaching the temperatures around $10^{9}$\,K, that led to a strong high-energy tail in the spectrum and a switch of the beaming pattern from limb darkening at lower energies to a strong limb brightening at highest energies.
In the opposite case of high contribution of high-energy particles, a rather hot skin could still be produced, although at a lower temperature, but the spectrum and the beaming pattern were very similar to those of the non-heated atmospheres.
We also found that the surface gravity of the NS affects the strength of the high-energy tail of the spectrum.

Large change in the emergent radiation spectrum and its beaming properties can have a significant impact on the inferred NS and magnetosphere parameters such as those obtained by the NICER team from RMPs using pulse profile modelling \citep{MLD_nicer19,RWB_nicer19}. 
The deviations of the angular distribution of specific intensity from that predicted by standard deep-heating models may exceed tens of per cent even in the thermal part of the spectrum, when the high-energy tail is nearly invisible.  
These deviations produce potentially orders of magnitude larger changes to the pulse profiles than the systematic uncertainty assumed in NICER studies \citep{BLM_nicer19}.
The results of our work may be used to improve NS radius constraints using these data.
Finally, because the exact properties of the emergent radiation are strongly dependent on the magnetospheric return-current energy distribution, the presented new atmosphere models can, in the future, be used to probe the still unknown structure and pair-production physics of the pulsar magnetospheres.

\section*{Acknowledgments}

This research was supported by the University of Turku Graduate School in Physical and Chemical Sciences (TS), by the Ministry of Science and Higher Education of the Russian Federation grant 14.W03.31.0021 (JP, VFS), the Deutsche Forschungsgemeinschaft (DFG) grant WE 1312/51-1 (VFS), the German Academic Exchange Service (DAAD) travel grants 57405000 and 57525212 (VFS), and the Academy of Finland grants 317552, 331951 and 333112 (TS, JP).
TS thanks Cole Miller and Ilia Kosenkov for useful discussions.
The computer resources of the Finnish IT Center for Science (CSC) and the Finnish Grid and Cloud Infrastructure project are acknowledged.

\bibliographystyle{aa}
\bibliography{allbib}

\begin{thebibliography}{62}
\expandafter\ifx\csname natexlab\endcsname\relax\def\natexlab#1{#1}\fi

\bibitem[{{Alcock} \& {Illarionov}(1980)}]{AI1980}
{Alcock}, C. \& {Illarionov}, A. 1980, \apj, 235, 534

\bibitem[{{Alme} \& {Wilson}(1973)}]{AW73}
{Alme}, M.~L. \& {Wilson}, J.~R. 1973, \apj, 186, 1015

\bibitem[{{Alpar} {et~al.}(1982){Alpar}, {Cheng}, {Ruderman}, \&
  {Shaham}}]{ACR82}
{Alpar}, M.~A., {Cheng}, A.~F., {Ruderman}, M.~A., \& {Shaham}, J. 1982, \nat,
  300, 728

\bibitem[{{Arons}(1981)}]{arons81}
{Arons}, J. 1981, \apj, 248, 1099

\bibitem[{{Baub{\"o}ck} {et~al.}(2019){Baub{\"o}ck}, {Psaltis}, \&
  {{\"O}zel}}]{BPO19}
{Baub{\"o}ck}, M., {Psaltis}, D., \& {{\"O}zel}, F. 2019, \apj, 872, 162

\bibitem[{{Beloborodov}(2008)}]{Belo08}
{Beloborodov}, A.~M. 2008, \apjl, 683, L41

\bibitem[{{Beloborodov} \& {Thompson}(2007)}]{BT07}
{Beloborodov}, A.~M. \& {Thompson}, C. 2007, \apj, 657, 967

\bibitem[{{Berger} {et~al.}(1984){Berger}, {Inokuti}, {Anderson}, {Bichsel},
  {Dennis}, {Powers}, {Seltzer}, \& {Turner}}]{Berger1984}
{Berger}, M.~J., {Inokuti}, M., {Anderson}, H.~H., {et~al.} 1984, Journal ICRU,
  os19, 1

\bibitem[{{Beskin}(2018)}]{Beskin2018}
{Beskin}, V.~S. 2018, Physics Uspekhi, 61, 353

\bibitem[{{Bogdanov}(2013)}]{bogdanov2013}
{Bogdanov}, S. 2013, \apj, 762, 96

\bibitem[{{Bogdanov}(2016)}]{bogdanov2016}
{Bogdanov}, S. 2016, European Physical Journal A, 52, 37

\bibitem[{{Bogdanov} {et~al.}(2019){Bogdanov}, {Lamb}, {Mahmoodifar}, {Miller},
  {Morsink}, {Riley}, {Strohmayer}, {Tung}, {Watts}, {Dittmann}, {Chakrabarty},
  {Guillot}, {Arzoumanian}, \& {Gendreau}}]{BLM_nicer19}
{Bogdanov}, S., {Lamb}, F.~K., {Mahmoodifar}, S., {et~al.} 2019, \apjl, 887,
  L26

\bibitem[{{Brambilla} {et~al.}(2018){Brambilla}, {Kalapotharakos}, {Timokhin},
  {Harding}, \& {Kazanas}}]{BKT2018}
{Brambilla}, G., {Kalapotharakos}, C., {Timokhin}, A.~N., {Harding}, A.~K., \&
  {Kazanas}, D. 2018, \apj, 858, 81

\bibitem[{{Cerutti} \& {Beloborodov}(2017)}]{Cerutti2017}
{Cerutti}, B. \& {Beloborodov}, A.~M. 2017, \ssr, 207, 111

\bibitem[{{Cerutti} {et~al.}(2016){Cerutti}, {Philippov}, \&
  {Spitkovsky}}]{CPS2016}
{Cerutti}, B., {Philippov}, A.~A., \& {Spitkovsky}, A. 2016, \mnras, 457, 2401

\bibitem[{{Chen} \& {Beloborodov}(2014)}]{Chen2014}
{Chen}, A.~Y. \& {Beloborodov}, A.~M. 2014, \apjl, 795, L22

\bibitem[{{Deufel} {et~al.}(2001){Deufel}, {Dullemond}, \& {Spruit}}]{DDS2001}
{Deufel}, B., {Dullemond}, C.~P., \& {Spruit}, H.~C. 2001, \aap, 377, 955

\bibitem[{{Gendreau} {et~al.}(2016){Gendreau}, {Arzoumanian}, {Adkins},
  {Albert}, {Anders}, {Aylward}, {Baker}, {Balsamo}, {Bamford}, {Benegalrao},
  {Berry}, {Bhalwani}, {Black}, {Blaurock}, {Bronke}, {Brown}, {Budinoff},
  {Cantwell}, {Cazeau}, {Chen}, {Clement}, {Colangelo}, {Coleman},
  {Coopersmith}, {Dehaven}, {Doty}, {Egan}, {Enoto}, {Fan}, {Ferro}, {Foster},
  {Galassi}, {Gallo}, {Green}, {Grosh}, {Ha}, {Hasouneh}, {Heefner}, {Hestnes},
  {Hoge}, {Jacobs}, {J{\o}rgensen}, {Kaiser}, {Kellogg}, {Kenyon}, {Koenecke},
  {Kozon}, {LaMarr}, {Lambertson}, {Larson}, {Lentine}, {Lewis}, {Lilly},
  {Liu}, {Malonis}, {Manthripragada}, {Markwardt}, {Matonak}, {Mcginnis},
  {Miller}, {Mitchell}, {Mitchell}, {Mohammed}, {Monroe}, {Montt de Garcia},
  {Mul{\'e}}, {Nagao}, {Ngo}, {Norris}, {Norwood}, {Novotka}, {Okajima},
  {Olsen}, {Onyeachu}, {Orosco}, {Peterson}, {Pevear}, {Pham}, {Pollard},
  {Pope}, {Powers}, {Powers}, {Price}, {Prigozhin}, {Ramirez}, {Reid},
  {Remillard}, {Rogstad}, {Rosecrans}, {Rowe}, {Sager}, {Sanders}, {Savadkin},
  {Saylor}, {Schaeffer}, {Schweiss}, {Semper}, {Serlemitsos}, {Shackelford},
  {Soong}, {Struebel}, {Vezie}, {Villasenor}, {Winternitz}, {Wofford},
  {Wright}, {Yang}, \& {Yu}}]{nicer2016}
{Gendreau}, K.~C., {Arzoumanian}, Z., {Adkins}, P.~W., {et~al.} 2016, in
  Society of Photo-Optical Instrumentation Engineers (SPIE) Conference Series,
  Vol. 9905, \procspie, 99051H

\bibitem[{{Goldreich} \& {Julian}(1969)}]{GJ69}
{Goldreich}, P. \& {Julian}, W.~H. 1969, \apj, 157, 869

\bibitem[{{Gonz{\'a}lez-Caniulef} {et~al.}(2019){Gonz{\'a}lez-Caniulef},
  {Zane}, {Turolla}, \& {Wu}}]{GCZT19}
{Gonz{\'a}lez-Caniulef}, D., {Zane}, S., {Turolla}, R., \& {Wu}, K. 2019,
  \mnras, 483, 599

\bibitem[{{Gould}(1972)}]{Gould1972}
{Gould}, R.~J. 1972, Physica, 60, 145

\bibitem[{{Guillot} {et~al.}(2019){Guillot}, {Kerr}, {Ray}, {Bogdanov},
  {Ransom}, {Deneva}, {Arzoumanian}, {Bult}, {Chakrabarty}, {Gendreau}, {Ho},
  {Jaisawal}, {Malacaria}, {Miller}, {Strohmayer}, {Wolff}, {Wood}, {Webb},
  {Guillemot}, {Cognard}, \& {Theureau}}]{GKR_nicer19}
{Guillot}, S., {Kerr}, M., {Ray}, P.~S., {et~al.} 2019, \apjl, 887, L27

\bibitem[{{Haakonsen} {et~al.}(2012){Haakonsen}, {Turner}, {Tacik}, \&
  {Rutledge}}]{mcphac2012}
{Haakonsen}, C.~B., {Turner}, M.~L., {Tacik}, N.~A., \& {Rutledge}, R.~E. 2012,
  \apj, 749, 52

\bibitem[{{Harding} \& {Muslimov}(2001)}]{HM2001}
{Harding}, A.~K. \& {Muslimov}, A.~G. 2001, \apj, 556, 987

\bibitem[{{Harding} \& {Muslimov}(2002)}]{HM02}
{Harding}, A.~K. \& {Muslimov}, A.~G. 2002, \apj, 568, 862

\bibitem[{{Haug}(2004)}]{H04}
{Haug}, E. 2004, \aap, 423, 793

\bibitem[{{Heinke} {et~al.}(2006){Heinke}, {Rybicki}, {Narayan}, \&
  {Grindlay}}]{HRN06}
{Heinke}, C.~O., {Rybicki}, G.~B., {Narayan}, R., \& {Grindlay}, J.~E. 2006,
  \apj, 644, 1090

\bibitem[{{Heitler}(1954)}]{Heitler1954}
{Heitler}, W. 1954, {Quantum theory of radiation}, International Series of
  Monographs on Physics (Oxford: Clarendon)

\bibitem[{{Ho} \& {Heinke}(2009)}]{HH09}
{Ho}, W.~C.~G. \& {Heinke}, C.~O. 2009, \nat, 462, 71

\bibitem[{{Ho} {et~al.}(2007){Ho}, {Kaplan}, {Chang}, {van Adelsberg}, \&
  {Potekhin}}]{Ho2007}
{Ho}, W. C.~G., {Kaplan}, D.~L., {Chang}, P., {van Adelsberg}, M., \&
  {Potekhin}, A.~Y. 2007, \mnras, 375, 821

\bibitem[{{Ibragimov} \& {Poutanen}(2009)}]{IP09}
{Ibragimov}, A. \& {Poutanen}, J. 2009, \mnras, 400, 492

\bibitem[{{Kurucz}(1970)}]{K70}
{Kurucz}, R.~L. 1970, SAO Special Report, 309

\bibitem[{{Lattimer}(2012)}]{lattimer12}
{Lattimer}, J.~M. 2012, Annual Review of Nuclear and Particle Science, 62, 485

\bibitem[{{Miller} \& {Lamb}(2015)}]{ML15}
{Miller}, M.~C. \& {Lamb}, F.~K. 2015, \apj, 808, 31

\bibitem[{{Miller} {et~al.}(2019){Miller}, {Lamb}, {Dittmann}, {Bogdanov},
  {Arzoumanian}, {Gendreau}, {Guillot}, {Harding}, {Ho}, {Lattimer}, {Ludlam},
  {Mahmoodifar}, {Morsink}, {Ray}, {Strohmayer}, {Wood}, {Enoto}, {Foster},
  {Okajima}, {Prigozhin}, \& {Soong}}]{MLD_nicer19}
{Miller}, M.~C., {Lamb}, F.~K., {Dittmann}, A.~J., {et~al.} 2019, \apjl, 887,
  L24

\bibitem[{{Morsink} {et~al.}(2007){Morsink}, {Leahy}, {Cadeau}, \&
  {Braga}}]{MLC07}
{Morsink}, S.~M., {Leahy}, D.~A., {Cadeau}, C., \& {Braga}, J. 2007, \apj, 663,
  1244

\bibitem[{{N{\"a}ttil{\"a}} \& {Pihajoki}(2018)}]{NP18}
{N{\"a}ttil{\"a}}, J. \& {Pihajoki}, P. 2018, \aap, 615, A50

\bibitem[{{N{\"a}ttil{\"a}} {et~al.}(2015){N{\"a}ttil{\"a}}, {Suleimanov},
  {Kajava}, \& {Poutanen}}]{NSK15}
{N{\"a}ttil{\"a}}, J., {Suleimanov}, V.~F., {Kajava}, J.~J.~E., \& {Poutanen},
  J. 2015, \aap, 581, A83

\bibitem[{{Olson} \& {Kunasz}(1987)}]{OK87}
{Olson}, G.~L. \& {Kunasz}, P.~B. 1987, \jqsrt, 38, 325

\bibitem[{{{\"O}zel} \& {Freire}(2016)}]{OF16}
{{\"O}zel}, F. \& {Freire}, P. 2016, \araa, 54, 401

\bibitem[{{Philippov} \& {Spitkovsky}(2018)}]{Philippov2018}
{Philippov}, A.~A. \& {Spitkovsky}, A. 2018, \apj, 855, 94

\bibitem[{{Poutanen} \& {Beloborodov}(2006)}]{PB06}
{Poutanen}, J. \& {Beloborodov}, A.~M. 2006, \mnras, 373, 836

\bibitem[{{Poutanen} \& {Gierli{\'n}ski}(2003)}]{PG03}
{Poutanen}, J. \& {Gierli{\'n}ski}, M. 2003, \mnras, 343, 1301

\bibitem[{{Radhakrishnan} \& {Srinivasan}(1982)}]{RS82}
{Radhakrishnan}, V. \& {Srinivasan}, G. 1982, Current Science, 51, 1096

\bibitem[{{Riley} {et~al.}(2019){Riley}, {Watts}, {Bogdanov}, {Ray}, {Ludlam},
  {Guillot}, {Arzoumanian}, {Baker}, {Bilous}, {Chakrabarty}, {Gendreau},
  {Harding}, {Ho}, {Lattimer}, {Morsink}, \& {Strohmayer}}]{RWB_nicer19}
{Riley}, T.~E., {Watts}, A.~L., {Bogdanov}, S., {et~al.} 2019, \apjl, 887, L21

\bibitem[{{Ruderman} \& {Sutherland}(1975)}]{RS75}
{Ruderman}, M.~A. \& {Sutherland}, P.~G. 1975, \apj, 196, 51

\bibitem[{{Salmi} {et~al.}(2018){Salmi}, {N{\"a}ttil{\"a}}, \&
  {Poutanen}}]{SNP18}
{Salmi}, T., {N{\"a}ttil{\"a}}, J., \& {Poutanen}, J. 2018, \aap, 618, A161

\bibitem[{{Salmi} {et~al.}(2019){Salmi}, {Suleimanov}, \& {Poutanen}}]{SSP19}
{Salmi}, T., {Suleimanov}, V.~F., \& {Poutanen}, J. 2019, \aap, 627, A39

\bibitem[{{Solodov} \& {Betti}(2008)}]{SB2008}
{Solodov}, A.~A. \& {Betti}, R. 2008, Physics of Plasmas, 15, 042707

\bibitem[{{Suleimanov} {et~al.}(2011){Suleimanov}, {Poutanen}, \&
  {Werner}}]{SPW11}
{Suleimanov}, V., {Poutanen}, J., \& {Werner}, K. 2011, \aap, 527, A139

\bibitem[{{Suleimanov} {et~al.}(2012){Suleimanov}, {Poutanen}, \&
  {Werner}}]{SPW12}
{Suleimanov}, V., {Poutanen}, J., \& {Werner}, K. 2012, \aap, 545, A120

\bibitem[{{Suleimanov} {et~al.}(2018){Suleimanov}, {Poutanen}, \&
  {Werner}}]{SPW18}
{Suleimanov}, V.~F., {Poutanen}, J., \& {Werner}, K. 2018, \aap, 619, A114

\bibitem[{{Timokhin}(2010)}]{T10}
{Timokhin}, A.~N. 2010, \mnras, 408, 2092

\bibitem[{{Timokhin} \& {Arons}(2013)}]{TA13}
{Timokhin}, A.~N. \& {Arons}, J. 2013, \mnras, 429, 20

\bibitem[{{van Kerkwijk} \& {Kaplan}(2007)}]{vkk07}
{van Kerkwijk}, M.~H. \& {Kaplan}, D.~L. 2007, \apss, 308, 191

\bibitem[{{Watts} {et~al.}(2016){Watts}, {Andersson}, {Chakrabarty}, {Feroci},
  {Hebeler}, {Israel}, {Lamb}, {Miller}, {Morsink}, {{\"O}zel}, {Patruno},
  {Poutanen}, {Psaltis}, {Schwenk}, {Steiner}, {Stella}, {Tolos}, \& {van der
  Klis}}]{WAC16}
{Watts}, A.~L., {Andersson}, N., {Chakrabarty}, D., {et~al.} 2016, Reviews of
  Modern Physics, 88, 021001

\bibitem[{{Watts} {et~al.}(2019){Watts}, {Yu}, {Poutanen}, {Zhang},
  {Bhattacharyya}, {Bogdanov}, {Ji}, {Patruno}, {Riley}, {Bakala}, {Baykal},
  {Bernardini}, {Bombaci}, {Brown}, {Cavecchi}, {Chakrabarty}, {Chenevez},
  {Degenaar}, {Del Santo}, {Di Salvo}, {Doroshenko}, {Falanga}, {Ferdman},
  {Feroci}, {Gambino}, {Ge}, {Greif}, {Guillot}, {Gungor}, {Hartmann},
  {Hebeler}, {Heger}, {Homan}, {Iaria}, {Zand}, {Kargaltsev}, {Kurkela}, {Lai},
  {Li}, {Li}, {Li}, {Linares}, {Lu}, {Mahmoodifar}, {M{\'e}ndez}, {Coleman
  Miller}, {Morsink}, {N{\"a}ttil{\"a}}, {Possenti}, {Prescod-Weinstein}, {Qu},
  {Riggio}, {Salmi}, {Sanna}, {Santangelo}, {Schatz}, {Schwenk}, {Song}, {{\v
  S}r{\'a}mkov{\'a}}, {Stappers}, {Stiele}, {Strohmayer}, {Tews}, {Tolos},
  {T{\"o}r{\"o}k}, {Tsang}, {Urbanec}, {Vacchi}, {Xu}, {Xu}, {Zane}, {Zhang},
  {Zhang}, {Zhang}, {Zheng}, \& {Zhou}}]{WYP19}
{Watts}, A.~L., {Yu}, W., {Poutanen}, J., {et~al.} 2019, Science China Physics,
  Mechanics, and Astronomy, 62, 29503

\bibitem[{{Zampieri} {et~al.}(1995){Zampieri}, {Turolla}, {Zane}, \&
  {Treves}}]{ZTZ95}
{Zampieri}, L., {Turolla}, R., {Zane}, S., \& {Treves}, A. 1995, \apj, 439, 849

\bibitem[{{Zane} {et~al.}(1998){Zane}, {Turolla}, \& {Treves}}]{ztt98}
{Zane}, S., {Turolla}, R., \& {Treves}, A. 1998, \apj, 501, 258

\bibitem[{{Zavlin} \& {Pavlov}(2002)}]{ZP2002}
{Zavlin}, V.~E. \& {Pavlov}, G.~G. 2002, in Neutron Stars, Pulsars, and
  Supernova Remnants, ed. W.~{Becker}, H.~{Lesch}, \& J.~{Tr{\"u}mper}, 263

\bibitem[{{Zavlin} {et~al.}(1996){Zavlin}, {Pavlov}, \& {Shibanov}}]{ZPS96}
{Zavlin}, V.~E., {Pavlov}, G.~G., \& {Shibanov}, Y.~A. 1996, \aap, 315, 141

\bibitem[{{Zel'dovich} \& {Shakura}(1969)}]{ZShak69}
{Zel'dovich}, Y.~B. \& {Shakura}, N.~I. 1969, \sovast, 13, 175

\end{thebibliography}

\end{document}